\documentclass[sn-nature, Numbered]{sn-jnl-modified}
\newcommand{\apjl}{Astrophys. J. Lett.}   
\newcommand{\apjs}{Astrophys. J. Suppl. Ser.}   
\newcommand{\ao}{Appl. Opt.}   
\newcommand{\aap}{Astron. Astrophys.}   

\newcommand{\vpsr}{PSR~B0833$-$45\xspace}
\newcommand{\hess}{H.E.S.S.\xspace}

\newcommand{\fermi}{{\it{Fermi}}\xspace}
\newcommand{\gar}{gamma-ray\xspace}
\newcommand{\gr}{gamma ray\xspace}
\newcommand{\grs}{\gr{}s\xspace}
\def\arcmin{\hbox{$^\prime$}}
\def\arcsec{\hbox{$^{\prime\prime}$}}

\usepackage{graphicx}%
\usepackage{multirow}%
\usepackage{amsmath,amssymb,amsfonts}%
\usepackage{amsthm}%
\usepackage{mathrsfs}%
\usepackage[title]{appendix}%
\usepackage{xcolor}%
\usepackage{textcomp}%
\usepackage{manyfoot}%
\usepackage{booktabs}%
\usepackage{algorithm}%
\usepackage{algorithmicx}%
\usepackage{algpseudocode}%
\usepackage{listings}%

\usepackage{xspace}
\usepackage{aas_macros}

\begin{document}

\title[Radiation reaching 20 Teraelectronvolt from the Vela Pulsar]{Discovery of a Radiation Component from the Vela Pulsar Reaching 20 Teraelectronvolts}

\author[]{
The \hess Collaboration:
F.~Aharonian$^{1,2}$,
F.~Ait~Benkhali$^{3}$,
J.~Aschersleben$^{4}$,
H.~Ashkar$^{5}$,
M.~Backes$^{6,7}$,
V.~Barbosa~Martins$^{8}$,
R.~Batzofin$^{9}$,
Y.~Becherini$^{10,11}$,
D.~Berge$^{8,12}$,
K.~Bernl\"ohr$^{2}$
% ...
\\
\\
\emph{
1. Dublin Institute for Advanced Studies, 31 Fitzwilliam Place, Dublin 2, Ireland \\
2. Max-Planck-Institut f\"ur Kernphysik, P.O. Box 103980, D 69029 Heidelberg, Germany \\
3. Landessternwarte, Universit\"at Heidelberg, K\"onigstuhl, D 69117 Heidelberg, Germany \\
4. GRAPPA, Anton Pannekoek Institute for Astronomy, University of Amsterdam,  Science Park 904, 1098 XH Amsterdam, The Netherlands \\
5. Laboratoire Leprince-Ringuet, École Polytechnique, CNRS, Institut Polytechnique de Paris, F-91128 Palaiseau, France \\
6. University of Namibia, Department of Physics, Private Bag 13301, Windhoek 10005, Namibia \\
7. Centre for Space Research, North-West University, Potchefstroom 2520, South Africa \\
8. DESY, D-15738 Zeuthen, Germany \\
9. School of Physics, University of the Witwatersrand, 1 Jan Smuts Avenue, Braamfontein, Johannesburg, 2050 South Africa \\
10. Université de Paris, CNRS, Astroparticule et Cosmologie, F-75013 Paris, France \\
11. Department of Physics and Electrical Engineering, Linnaeus University,  351 95 V\"axj\"o, Sweden \\
12. Institut f\"ur Physik, Humboldt-Universit\"at zu Berlin, Newtonstr. 15, D 12489 Berlin, Germany \\
13. Institut f\"ur Astronomie und Astrophysik, Universit\"at T\"ubingen, Sand 1, D 72076 T\"ubingen, Germany \\
14. Laboratoire Univers et Théories, Observatoire de Paris, Université PSL, CNRS, Université de Paris, 92190 Meudon, France \\
15. Sorbonne Universit\'e, Universit\'e Paris Diderot, Sorbonne Paris Cit\'e, CNRS/IN2P3, Laboratoire de Physique Nucl\'eaire et de Hautes Energies, LPNHE, 4 Place Jussieu, F-75252 Paris, France \\
16. Université Savoie Mont Blanc, CNRS, Laboratoire d'Annecy de Physique des Particules - IN2P3, 74000 Annecy, France \\
17. IRFU, CEA, Universit\'e Paris-Saclay, F-91191 Gif-sur-Yvette, France \\
18. Friedrich-Alexander-Universit\"at Erlangen-N\"urnberg, Erlangen Centre for Astroparticle Physics, Erwin-Rommel-Str. 1, D 91058 Erlangen, Germany \\
19. Astronomical Observatory, The University of Warsaw, Al. Ujazdowskie 4, 00-478 Warsaw, Poland \\
20. Instytut Fizyki J\c{a}drowej PAN, ul. Radzikowskiego 152, 31-342 Krak{\'o}w, Poland \\
21. University of Oxford, Department of Physics, Denys Wilkinson Building, Keble Road, Oxford OX1 3RH, UK \\
22. Institut f\"ur Physik und Astronomie, Universit\"at Potsdam,  Karl-Liebknecht-Strasse 24/25, D 14476 Potsdam, Germany \\
23. Aix Marseille Universit\'e, CNRS/IN2P3, CPPM, Marseille, France \\
24. School of Physical Sciences, University of Adelaide, Adelaide 5005, Australia \\
25. Laboratoire Univers et Particules de Montpellier, Universit\'e Montpellier, CNRS/IN2P3,  CC 72, Place Eug\`ene Bataillon, F-34095 Montpellier Cedex 5, France \\
26. Universit\'e Bordeaux, CNRS, LP2I Bordeaux, UMR 5797, F-33170 Gradignan, France \\
27. Institut f\"ur Astro- und Teilchenphysik, Leopold-Franzens-Universit\"at Innsbruck, A-6020 Innsbruck, Austria \\
28. Universit\"at Hamburg, Institut f\"ur Experimentalphysik, Luruper Chaussee 149, D 22761 Hamburg, Germany \\
29. Obserwatorium Astronomiczne, Uniwersytet Jagiello{\'n}ski, ul. Orla 171, 30-244 Krak{\'o}w, Poland \\
30. Institute of Astronomy, Faculty of Physics, Astronomy and Informatics, Nicolaus Copernicus University,  Grudziadzka 5, 87-100 Torun, Poland \\
31. Nicolaus Copernicus Astronomical Center, Polish Academy of Sciences, ul. Bartycka 18, 00-716 Warsaw, Poland \\
32. Department of Physics and Astronomy, The University of Leicester, University Road, Leicester, LE1 7RH, United Kingdom \\
33. Yerevan Physics Institute, 2 Alikhanian Brothers St., 375036 Yerevan, Armenia \\
34. Department of Physics, Konan University, 8-9-1 Okamoto, Higashinada, Kobe, Hyogo 658-8501, Japan \\
35. Kavli Institute for the Physics and Mathematics of the Universe (WPI), The University of Tokyo Institutes for Advanced Study (UTIAS), The University of Tokyo, 5-1-5 Kashiwa-no-Ha, Kashiwa, Chiba, 277-8583, Japan \\
36. RIKEN, 2-1 Hirosawa, Wako, Saitama 351-0198, Japan \\
37. Cherenkov Telescope Array Observatory gGmbH, Via Gobetti, Bologna, Italy\\
38. Space Science Division, Naval Research Laboratory, Washington 20375-5352, USA\\
39. CSIRO Astronomy and Space Science, Australia Telescope National Facility, PO Box 76, Epping, NSW 1710, Australia\\
40. Centre for Astrophysics and Supercomputing, Swinburne University of Technology Mail H30, PO Box 218, Hawthorn, VIC 3122, Australia\\
41. ARC Centre of Excellence for Gravitational Wave Discovery (OzGrav), Australia\\
42. Centre d’Etudes Nucléaires de Bordeaux Gradignan, IN2P3/CNRS, Université de Bordeaux 1, BP120, F-33175 Gradignan Cedex, France\\
43. Laboratoire d’Astrophysique de Bordeaux, Université de Bordeaux, CNRS, B18N, allée Geoffroy Saint-Hilaire, F-33615 Pessac, France\\
}
\begin{flushleft}
{\raggedright Corresponding authors: \indent\,\,\, djannati@in2p3.fr, emma.de.ona.wilhelmi@desy.de, bronek@ncac.torun.pl, christo.venter@nwu.ac.za,
      %
      contact.hess@hess-experiment.eu} 
\end{flushleft}
}


    \abstract{
      Gamma-ray observations have established energetic isolated pulsars as outstanding
      particle accelerators and antimatter factories in the Galaxy.
      There is, however, no consensus regarding the acceleration mechanisms and the radiative
processes at play, nor the locations where these take place.
      The spectra of all observed gamma-ray pulsars to date show strong cutoffs or a break above energies of 
      a few gigaelectronvolt (GeV). Using the \hess{} array of Cherenkov telescopes,
      we discovered a novel radiation component emerging beyond this generic GeV cutoff 
      in the Vela pulsar's broadband spectrum. The extension of gamma-ray pulsation energies
      up to at least 20 teraelectronvolts (TeV) shows that Vela pulsar can accelerate particles to Lorentz
      factors higher than  $4\times10^{7}$.
      This is an order of magnitude larger than in the case of the Crab pulsar,
the only other pulsar detected in the TeV energy range.
Our results challenge the state-of-the-art models for high-energy emission of pulsars
while providing a new probe, i.e. the energetic multi-TeV component, for constraining the
acceleration and emission processes in their extreme energy limit.

}


\keywords{{gamma-rays: stars -- pulsars: individual: Vela pulsar (\vpsr) -- radiation mechanisms: non-thermal}}



\maketitle


Pulsars, the progeny of supernova explosions, are rapidly spinning
and strongly magnetized
neutron stars that emit beams of electromagnetic radiation modulated
at the stellar rotational period.
Their radiation spans a wide range of frequencies -- from the radio domain,
where more than 3,000 pulsars are known~\cite{atnf}, to high-energy (HE; 100~MeV$-$100~GeV) \grs{},
where the number of identified or discovered pulsars exceeds 270~\cite{Abdo20132PC}. 
Gamma rays are widely believed to be emitted by electrons and positrons (electrons, hereafter) accelerated
to TeV energies at the expense of the neutron star's
rotational energy. However, there is no consensus yet as to the origin of the
observed pulsed signals.

Gamma rays in the HE range have proven to be essential probes of pulsar magnetospheres.
Indeed, measurements of the spectra of bright gamma-ray pulsars by space-borne
telescopes ({\it EGRET}~\cite{VelaEgretKanbach94,EgretPulsarsNolan1996} and
\fermi-LAT~\cite{Abdo20132PC}) have established strong
(i.e. exponential) cutoffs at energies beyond a few GeV.
The cutoffs are not as abrupt as expected 
in magnetic photon absorption or photon splitting scenarios 
near the stellar poles, thus ruling out those regions as possible production sites for
GeV photons \cite{fermifirstvela,abdo2010}.
Very-high-energy (VHE; $> $100 GeV) \grs{} are
invaluable tools for testing acceleration and emission processes
in their extreme energy limit. They are, however, out of reach for
satellites but accessible to ground-based telescopes.  
Previous searches for pulsations in the VHE domain have resulted 
in the  detection up to an energy $\sim 1$~TeV of only one pulsar, that associated with the Crab nebula \cite{veritasvhecrab,Aleksic2012,Ansoldi2016},
while providing stringent upper limits on the VHE fluxes of other pulsars \cite{HessPSRULs2007,Veritas13psrs2019}.

We report here on the detection of the Vela pulsar (hereafter Vela),
\vpsr, in the multi-TeV energy range with the \hess array of five imaging atmospheric Cherenkov telescopes (CT1-5).
\hess is situated in the Khomas Highland of Namibia and operates in the tens
of GeV to tens of TeV energy range.  Vela was among the
very first pulsars discovered at radio frequencies \cite{VelaMolongo68}, the
second one detected in $>30$~MeV \grs \cite{Thompson1975}, and
stands out as being by far the brightest pulsar in these two domains.
Located nearby, at a distance of 287~pc\cite{Dodson2003}, it is a young
pulsar with a spin period of 89~ms and a characteristic age of 11~kyr.
In the GeV range, its rotation phase-folded \gar light curve exhibits two peaks, labeled P1
and P2, separated by 0.43 in phase and connected by a bridge emission containing a third peak
labeled P3 \cite{fermifirstvela,VelaMonoPaper}.
Recently, using the largest \hess telescope, CT5, 
which, thanks to its 28-m equivalent diameter provides a relatively low energy threshold, 
we detected the P2 pulse of Vela in the $10-80$~GeV energy range and showed that there
was compelling evidence that the bright GeV component has a cutoff at energies well below
100~GeV \cite{VelaMonoPaper}.

Results reported here are based on
deeper observations (80 hours) above an energy threshold of 260~GeV,
performed with the 12-m diameter CT1-4 telescopes during the 2004-2007 and 2014-2016 observing
seasons \cite{SOM}.
Given the lack of {\it a priori} knowledge of the source spectral hardness (whether soft or hard, i.e.
dominated by events with energy below $\sim$1~TeV, or {\it vice versa}),   
the search for pulsations was conducted 
by applying periodicity tests on data sets selected using
four predefined and increasing energy thresholds of 0.5, 1, 3 and 7 TeV.
Three types of periodicity tests were used: the H-test \cite{deJager89} where no \textit{a priori}
knowledge of the light curve (or phasogram, i.e., the phase-folded distribution of
events) is assumed, the C-test \cite{deJager1994} where the position and the
(approximate) width of the pulse shape are supposed to be known beforehand, and 
a maximum likelihood-ratio (LR) test \cite{LiAndMa83} based on \textit{a priori} defined
\textit{On}-- and \textit{Off}-phase intervals. 
The pulse P2 of the Vela pulsar, dominating in the
tens of GeV energy range, was considered as the prime candidate for detection in the VHE range
and its parameters, as derived from the \fermi-LAT phasogram above 10 GeV \cite{VelaMonoPaper}, were used
as input to the tests.
Pulsed emission was detected at a statistical significance exceeding $4\sigma$ 
for all the tests: above energy thresholds of 1, 3 and 7 TeV with the C-test (4.3, 4.9 and
5.6$\sigma$, respectively), 3 and 7 TeV with the LR test (4.7 and
4.8$\sigma$, respectively), and above 7 TeV for the H-test (4.5$\sigma$).
Posterior to this detection, we derived the significance of the pulsations above two other energy thresholds, 5 and 20 TeV.
The signal displays its highest significance level above 5 TeV and is clearly detected above 20 TeV, with, e.g.,
C-test results of 5.8 and 4.6$\sigma$, respectively (Table~S1).

\begin{figure*}
\hspace{-1cm} \centering
\includegraphics[width=10cm]{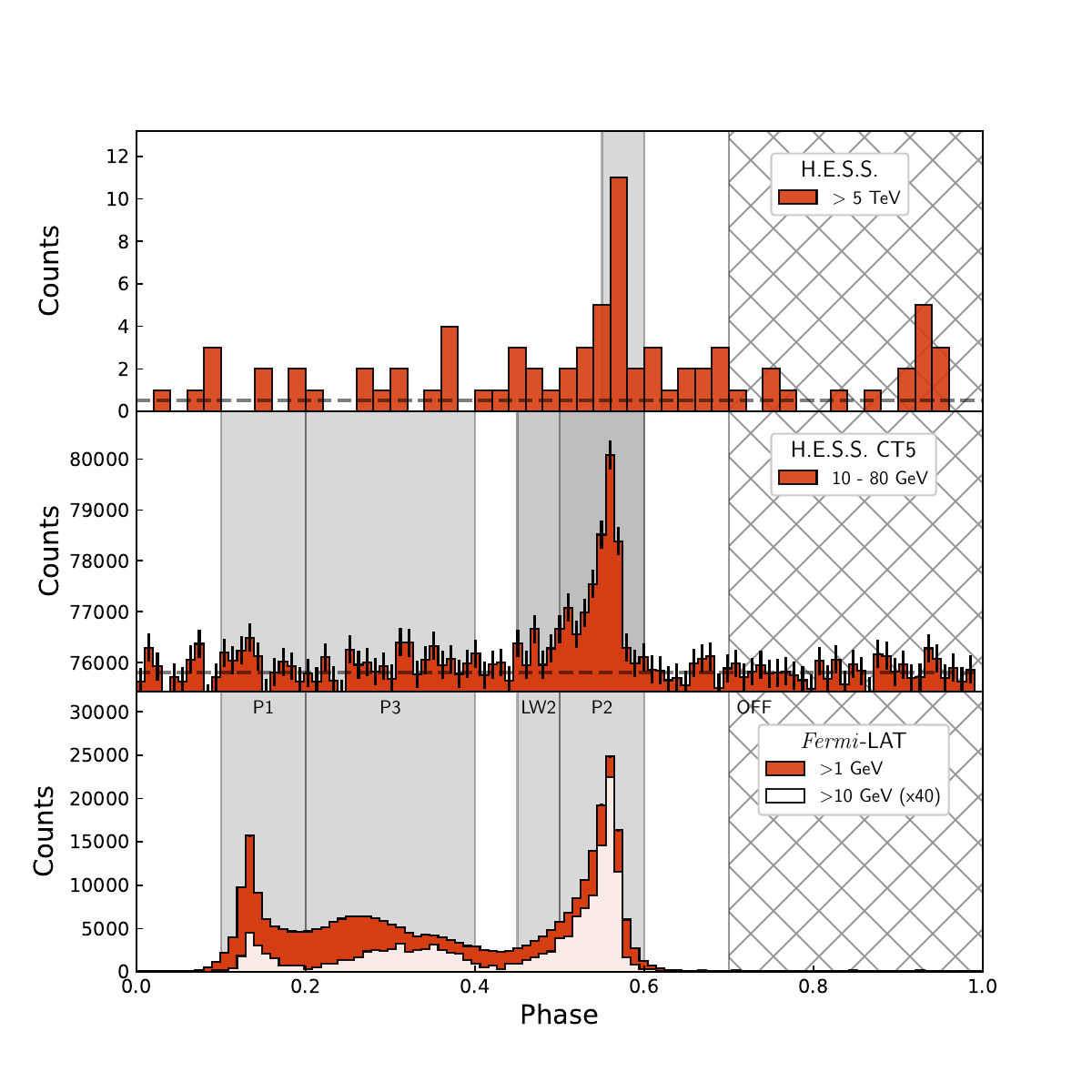}
\caption{
  {\bf Phasogram of Vela as measured with \hess CT1-4 for energies above 5~TeV, with \hess CT5 in the $10-80$~GeV range
    and with the \fermi-LAT above 1~GeV and 10~GeV, respectively.}
  Phase values are computed relative to the radio pulse. The ranges corresponding to different features in the pulse profile at low energies ($<$100~  GeV) are shown as grey-colored intervals: pulses P1, P3, P2, and the leading wing of P2, LW2.
  The {\it off}-phase interval $[0.7 - 1.0]$ is shown as a hatched area  and the dashed line on the
  two upper panels shows the estimated level of the background \cite{VelaMonoPaper}. 
  The \fermi-LAT light curve for energies above $10$~GeV has been multiplied by a factor 40 for better visibility.
}
\label{fig::phaso1}
\end{figure*}

Fig.~1 
shows the phasogram of Vela obtained with 
and the \fermi-LAT.
The sole significant feature present in the multi-TeV
range lies at a peak position ($\phi_{\rm P2}^{\rm TeV}=0.568\pm0.003 $) that is statistically
compatible with that of the P2 pulse observed in the HE  energy range
($\phi_{\rm P2}^{\rm GeV}=0.565\pm0.001 $). This pulsation also exhibits a similar width (e.g., FWHM) to that
measured in the latter energy range. 
This, together with the fact that P1 disappears above few tens of GeV, 
is in line with the energy evolution of the light curve at GeV energies \cite{SOM}.

\begin{figure*}
\centering
\includegraphics[width=10cm]{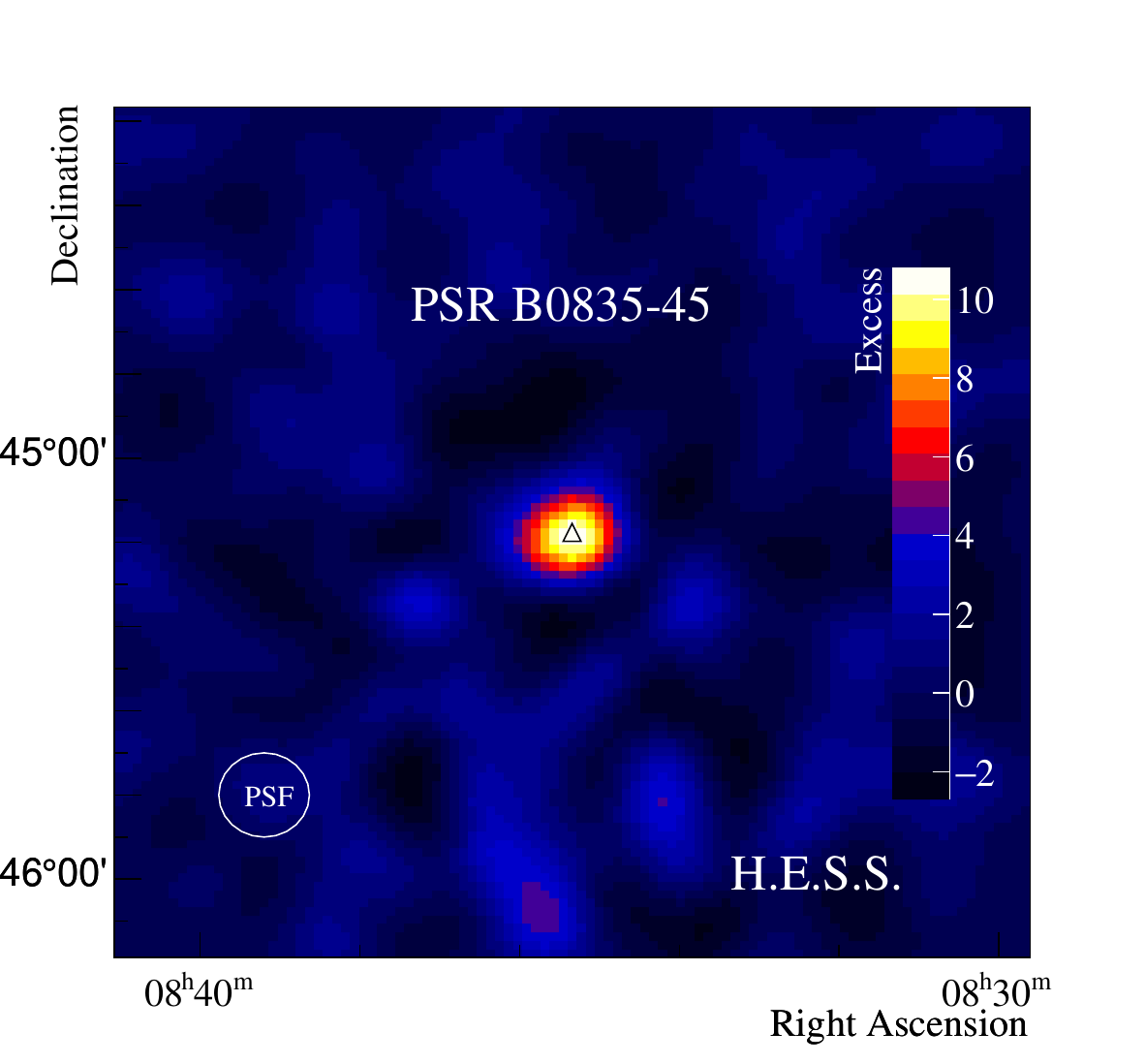}
\caption{
  {\bf Excess map of the P2 pulse of Vela as measured with \hess for energies above 5~TeV.}
  Gaussian-smoothed excess map ($\sigma=0.15^{\circ}$) in the P2 phase range, 
  where the {\it on} and {\it off} maps are made after selection of events in {\it on}- and {\it off}-phase intervals defined as $[0.55-0.6]$ and $[0.7-1.0]$, respectively.
  The triangle indicates the position of the pulsar and the circle shows the 1$\sigma$ instrument point spread function.
}
\label{fig::pulsedmap}
\end{figure*}

\vspace{0.5cm} Fig.~2
shows the map of
photons with energies above 5~TeV in the P2 pulse phase range. 
The spatial distribution of photons is centered on Vela and its spread is
compatible with the \hess{} point-spread function, 
as expected for a point-like source.  
The spectral energy distribution of the P2 pulse is shown in Fig.~3. 
It was measured by
selecting signal and background events in the phase ranges of $[0.55 - 0.6]$ and $[0.7 - 1.0]$, respectively.  
The fit of a power-law function (${\rm d}N(E)/{\rm d}E = \Phi_0 \left({E}/{E_0}\right)^{-\Gamma_{\rm VHE}}$) in the
260 GeV $-$ 28.5 TeV energy range resulted in a very hard spectrum
with photon index $\Gamma_{\rm VHE} = 1.4 \pm 0.3^{\rm stat} \pm
0.1^{\rm syst}$, and normalization 
$\Phi_0= \left(1.74 \pm 0.52^{\rm stat} \pm 0.35^{\rm syst}\right) \times 10^{-15}$ {erg}$^{-1}$cm$^{-2}$s$^{-1}$
at the 
reference energy $E_0 = 4.24 $~TeV,
implying an isotropic luminosity $L_{20 \rm TeV} \simeq 2 \times 10^{30} {\rm erg \, s^{-1}} $.
Given the steeply falling HE spectrum (photon index $\Gamma_{\rm HE} = 5.25 \pm 0.25^{\rm stat}$~\cite{VelaMonoPaper}), and the non detection upper limits in the 100-660~GeV range, the extremely hard VHE spectrum can only be a distinct and new component.

\begin{figure*}[hp]
  \centering
  \includegraphics[width=\linewidth]{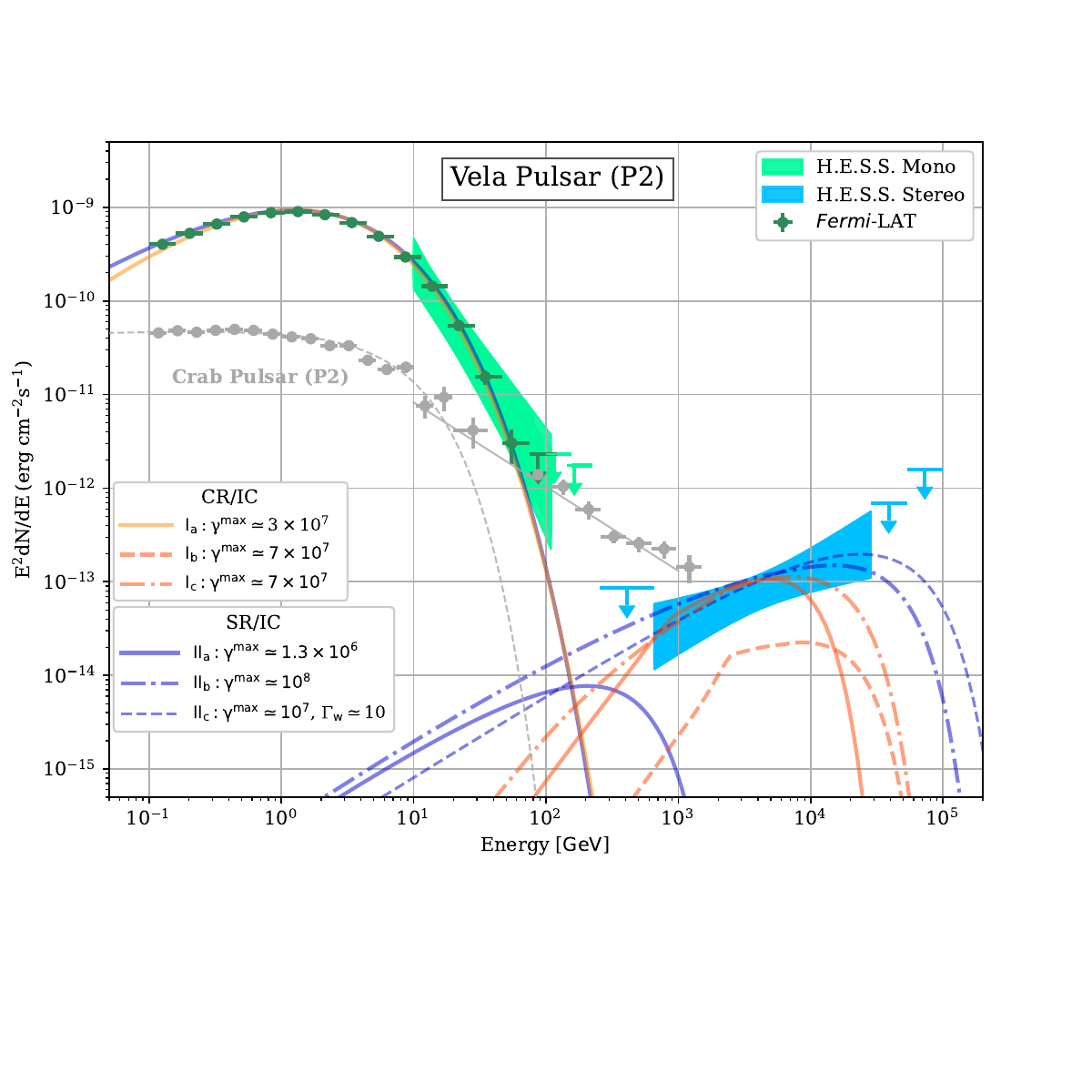}
\vspace{-2cm}
 \caption{\small
{\bf Spectral energy distribution (SED) of the P2 pulse of Vela }\newline
{\it Data}: The green points and the green area below 100 GeV   
show the measurements by \fermi-LAT and by H.E.S.S. CT5 in monoscopic mode \cite{VelaMonoPaper}, respectively.
The blue area and upper limits (ULs) above 260\,GeV correspond to measurements with
\hess{} CT1-4 in stereoscopic mode (this work).
All ULs are given at 99.7\% confidence level, and both 
CT5 and CT1-4 uncertainty bands consist of 1$\sigma$ confidence intervals combined
with systematic errors on the \hess{} energy scale.
For comparison, the SED of
the P2 pulse of the Crab pulsar as measured by \fermi-LAT and MAGIC
\cite{Ansoldi2016} is also shown.
\newline
    {\it Heuristic spectral models}: Either magnetospheric curvature radiation (CR) or synchrotron radiation (SR) in the wind zone 
    is considered for emission below 100 GeV, while for 
    the TeV range inverse-Compton (IC) scattering of soft photons is assumed (see illustration in Fig.~4 
    and \cite{SOM}).
    The CR/IC and SR/IC schemes are shown in
 orange 
 and blue colors, 
 respectively.  
 The \hess data require $\gamma^{\rm max} \gtrsim 7\times 10^7$ and hence
 exclude the traditional scenarios $\rm{I_a}$ (CR/IC), i.e.
 emission in the inner magnetosphere or at the light cylinder (LC), and $\rm{II_a}$ (SR/IC), where 
 $\gamma^{\rm max}$ is limited by SR cooling.
 The dashed and dash-dotted curves show possible paths to fit the data, including
 a Doppler-boosted scenario ($\rm{II_c}$) with bulk wind Lorentz factor $\Gamma_{\rm w} \simeq 10$  (see text). 
 The IC intensity is only loosely constrained due to its strong dependence 
 on model-dependent geometrical factors and on the density of putative target photons, which in turn depends on the
 unknown lower limit of its spectrum.
 All spectral models are computed with IC seed photons extending into the FIR domain ($0.005 - 4$~eV)
 and are normalized (for $\rm{II_a}$, its extrapolation) to a level of  $10^{-13}$ {erg}$^{-1}$cm$^{-2}$s$^{-1}$ at 5~TeV,
 except $\rm{I_c}$.
 For the latter model, the targets are limited to the O-NIR range ($0.1 - 4$~eV).
More sophisticated approaches in the CR/IC scenario are
shown in Fig.~\ref{fig::sedmodels}.
 }
\label{fig::spectral_models}
\end{figure*}

The most likely process for
producing \grs at multi-TeV energies by energetic electrons, whatever the
acceleration mechanism and emission regions are, is inverse-Compton
(IC) scattering of low-energy photons (e.g.
\cite{chr1986a,romani96, Hirotani2001,bogovalov2000}).
Potential target photon fields in Vela may include the observed non-thermal
X-rays~\cite{rxte2002,manzali2007}, thermal X-rays from the neutron
star surface \cite{manzali2007}, UV~\cite{uv2005}, or
optical~\cite{velaoptical2007,velaoptical2019} to the near-infrared 
\cite{zyu2013} emission (O-NIR).
For all these photons IC scattering would proceed in the Klein-Nishina (K-N) regime.  
The O-NIR radiation field with its possible extension down to the far-infrared (FIR) domain constitutes
the most plausible target \cite{chr1986a,romani96, Hirotani2001,ahaaronianbogovalov2003}.  
In the K-N regime, the maximum measured photon energy $E^{\rm max} \gtrsim 20\,{\rm TeV}$
constrains the electron Lorentz factor to be $\gamma_{\rm IC}^{\rm max} \gtrsim E^{\rm max}/{m_{\rm e} c}^2  \gtrsim 4\times 10^7$.
The fact that P2 in Vela
occurs at the same phase position for both spectral components, HE and
VHE, suggests that these components 
are generated by the same population of electrons, 
but through different radiation
processes.  
In the following we discuss the implications of the \hess{} discovery under this hypothesis. 

Several hypotheses have been proposed to describe the
acceleration of electrons to ultra-relativistic energies (see illustration in Fig.~4). 
In a first scenario, particles are accelerated along the magnetic field lines
in the pulsar magnetosphere by the (unscreened) electric field $E_{\parallel}$ that is
parallel to these local lines, in
charge-depleted cavities (or gaps) \cite{AronsScharlemann1979,chr1986a,muslimovh2003}
within the light cylinder 
(LC), or,  as recently posited, also slightly beyond \cite{ahvela2018}. 
The latter is defined as the radius at which the co-rotation speed equals that of light in vacuum
($R_{\rm LC} =cP/2\pi  \simeq4.3 \times 10^8\,\rm{cm}$ for the Vela pulsar given its period $P$=89.3~ms).
In a second scenario, acceleration  takes place through magnetic reconnection in the equatorial current sheet (CS) of the striped wind beyond the LC \cite{Michel1971,Coroniti1990,Michel1994,Lyubarsky1996, KirkLyubarsky2001,kirk2002,petri2012}).  

In the first scenario, curvature radiation (CR) 
is traditionally posited to explain the emission
observed in the GeV range, e.g. \cite{romani96, Takata2006, Hirotani2008p, Takata2016}, and
a combination of synchrotron (SR) and curvature (synchro-curvature radiation, SCR),  
to also reproduce the MeV to GeV spectral shape (e.g., \cite{vig2015b, ahvela2018}).
The maximum Lorentz factor of the electrons is limited by the magnitude of the accelerating electric field $E_\parallel$ 
in the gap -- or equivalently, the magnetic conversion efficiency $\eta=E_\parallel/B$ -- and by CR losses which
depend on the curvature radius $\rho_{\rm c}$ of particle trajectories. This limit
can be expressed as $\gamma_{\rm CR}^{\rm max} \propto {\rho_{\rm c}^{1/2}}\, \eta^{1/4}$ \cite{SOM}.
The  magnitude of $\eta$ depends on the particular version of the acceleration
gaps\footnote{And may vary with altitude above the pulsar surface.} 
with values usually assumed to be below 10\% at the LC \cite{Hirotani2006,ahvela2018}.
Hence, to achieve the maximum photon energy observed by \hess{}, $E^{\rm max}$,
large curvature radii $\rho_{\rm c}\gtrsim 4\times 10^8\,\rm{cm} \approx R_{\rm LC}$ are required \cite{SOM}.
Taking into account the HE and VHE spectra provides further constraints.
The HE spectral peak lying at $E^{\rm  peak}_{\rm HE}\simeq$1.5\,GeV \cite{fermifirstvela,abdo2010,VelaMonoPaper}
depends also on the combination of $\eta$ and $\rho_{\rm c}$
as  $E_{\rm HE}^{\rm peak} \propto \rho_{\rm c}^{1/2}\, \eta^{3/4} $.
Considering first an emission zone close to the LC, where $\rho_{\rm c} \sim R_{\rm LC}$,
and fitting the GeV component alone results in $\gamma_{\rm CR}^{\rm max} \simeq  3\times 10^7$ and $\eta\simeq0.02$
which is insufficient to reproduce the TeV data (curve $\rm{I_a}$ in Fig.~\ref{fig::spectral_models}).
A joint fit of both components requires $\gamma_{\rm IC}^{\rm max}\gtrsim 7\times 10^7$,
and by identifying  $\gamma_{\rm IC}^{\rm max}$ with  $\gamma_{\rm CR}^{\rm max}$
we obtain $\eta \ll 0.1$ and  $\rho_{\rm c}\gg \, R_{\rm LC}$ (curve $\rm{I_b}$ in Fig.~\ref{fig::spectral_models},
red curve in Fig.~\ref{fig::etaxigamma}). 
Hence, if the HE and VHE components are produced by the same electron population,
the \hess{} data constrain the emission regions to lie beyond the LC and imply at the same time a low magnetic conversion efficiency.

In the second scenario, SR has been proposed as the mechanism responsible for
the GeV radiation
\cite{Lyubarsky1996,petri2012,arkadubus2013,cerutti2017}, and applied
to model the HE component of the Crab and Vela pulsars
\cite{mochol15,mochol17}.
Hard particle spectra reaching maximum energies beyond the radiative
cooling limit are expected in the magnetic reconnection scheme
\cite{kirk2004}, due to a two-step process: the acceleration takes place 
deep in the CS where the magnitude of the perpendicular $B$ is weak,
and is followed by abrupt SR cooling in the magnetic loops (plasmoids) where B-field is strong (e.g.,
\cite{Uzdensky2011, werner2016,cps2016, werner2017,cerutti2020}).
The sharp HE cutoffs observed at a few GeV in the spectra of pulsars are attributed
to the latter step. 
In the Vela case, the SR cutoff would correspond to a
maximum Lorentz factor of $\gamma_{\rm SR}^{\rm max}\simeq 1.3 \times 10^6$ \cite{SOM}.
The matching inferred particle spectral indices in the sub-GeV and TeV
regimes, and the compatible luminosity levels when considering the
available photon fields \cite{SOM}, renders the SR/IC scenario in the
dissipation region near the LC \cite{cerutti2020} attractive. However,
$\gamma_{\rm SR}^{\rm max}$ is two orders of magnitude lower than the one
derived from the \hess{} data  (curve $\rm{II_a}$ in Fig.~3) 
and requires a more complex approach.
One can speculate on the escape of the highest energy (and IC-emitting)
particles from plasmoids, or their re-energization after SR cooling
\cite{Petropoulou2018,Hakobyan2019, cerutti2020}, or alternatively, assume
that two populations of electrons are responsible for the HE and VHE
components \cite{SOM}  (curve $\rm{II_b}$ in Fig.~3). 
Invoking a Doppler-boosted plasma as the origin of the
GeV and/or multi-TeV emission \cite{bogovalov2000,Aharonian2012,arkadubus2013,mochol15,mochol17,Tavernier2015, Jacob2019} alleviates
the tensions related to the maximum achievable energy in the SR/IC
scheme.  The \fermi-LAT and \hess{} data can be used to
constrain the wind Lorentz factor to be $\Gamma_{\rm w}\gtrsim 5$ at a
distance of $\simeq 5 R_{\rm LC}$ where the gamma-ray emission region
should be located  (curve $\rm{II_b}$ in Fig.~3).
This region is, however, further than the
typical zone at $\sim 1-2 R_{\rm LC}$ where the dissipation of the
energy is believed to occur through SR according to current Particle-in-cell (PIC)
simulations \cite{cerutti2020}. Resorting to differentiated SR and IC
cooling zones could mitigate this issue with the condition that the
photons from these zones are beamed into similar phases.

\begin{figure*}[tbh]
\centering
\includegraphics[width=\linewidth]{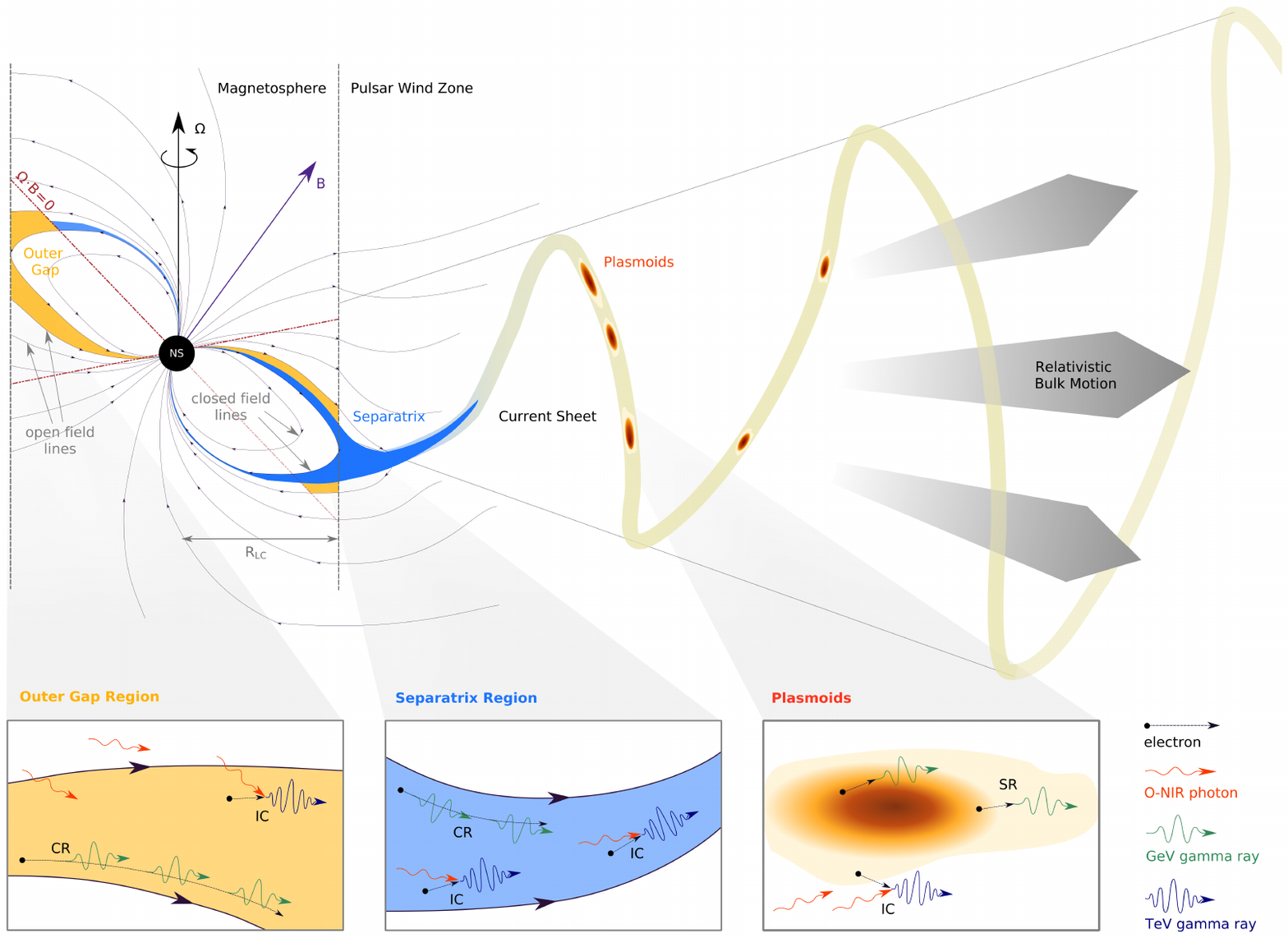}
\caption{
  {\bf Sketch illustrating main scenarios of particle acceleration and gamma-ray emission.}
  Electrons are accelerated either (i) along magnetic field lines in
  charge-depleted cavities within the light cylinder (LC), i.e. outer
  gaps, or slightly beyond, i.e. the separatix/current sheet model, or
  (ii) through magnetic reconnection in the equatorial current sheet of
  the striped wind beyond the LC.
  GeV gamma rays are either due to curvature radiation (CR) or synchrotron radiation (SR), while TeV photons
  are produced through inverse-Compton (IC) scattering of low-energy (O-NIR) photons (see text). 
  For sake of readability scales are not respected: the pulsar
  size is exaggerated as well as the size of the acceleration and emission zones.
  The neutron star (NS) has a diameter of $\sim12$~km and the light-cylinder radius $R_{\rm LC}\simeq 4300$~km. The wavelength of the
  current sheet stripes ($ 2 \times \pi \times R_{\rm LC} $) is twice as large as that depicted in the sketch.}
\label{fig::sketch}
\end{figure*}

Reproducing the HE-VHE light curves poses indeed further challenges. As
mentioned above, the TeV light curve maintains the trend observed
below 80~GeV, where the ratio of the intensities of the two peaks P1
and P2 decreases with energy (see Fig.~1 
and \cite{abdo2010,Abdo20132PC}), that is, the cutoff energy of the P2
spectrum is higher than that of P1.
To form the light curves as measured in the HE and VHE domains,
\gar{} photons should originate in radially extended and properly
shaped zones.  Special-relativistic effects and the $B$-field
structure arrange the photons to arrive at Earth at similar phases,
i.e., to form caustics.  Within (or slightly beyond) the
magnetosphere, these caustics arise naturally
\cite{morini1983,RY1995,crz2000,dr2003,dhr2004,BaiSpitkovsky2010a},
and the higher cutoff energies of the P2 spectral component can be
attributed to larger curvature radii of the orbits of electrons
responsible for P2 via CR \cite{ah2021,Barnard2022}. 
Caustics
can also form within the equatorial CS in the near wind zone
\cite{arkadubus2013,cps2016,Philippov2018}, resulting in double-peaked
light curves with specific predictions for the polarization of the HE
emission \cite{2016MNRAS.463L..89C}.  Alternatively, the phase
coherence of the pulsations can be obtained by a relativistic beaming
effect in the far wind zone \cite {Lyubarsky1996,
  KirkLyubarsky2001,kirk2002, petri2012}.  In this case, the higher
cutoff energy of the P2 pulse could arise from the difference in the
maximum energies attained by the positron and electron populations
with potentially distinct contributions to the pulses
\cite{cps2016,Philippov2018}.
However, as discussed above, both CR/IC and SR/IC scenarios are strongly challenged
by the \hess{} measurements.

\vspace{0.5cm}
The results reported here establish 
Vela as the first pulsating
source of tens of TeV gamma rays and as the second pulsar detected in the VHE range,
after the Crab pulsar \cite{Ansoldi2016}.
We find the dominant dissipation of energy, and thus particle acceleration and
photon emission, to happen beyond the pulsar LC or at its
periphery, and we set a lower limit of $4 \times 10^7\,m_e\,c^2 $ to the maximum achievable electron energy.
In contrast to the Crab pulsar, of which the hardest pulsation is
shown in Fig.~3, 
Vela unambiguously displays a new spectral component, with a very hard index, 
extending 
to energies an order of magnitude higher.
These are unprecedented challenges to the state-of-the-art models of HE and VHE emission from pulsars.

Our discovery opens a new observation window for detection of other pulsars
in the TeV to the tens of TeV range with current and upcoming more
sensitive instruments such as LHAASO~\cite{2020ChPhC..44f5001A} or CTA~\cite{2013CTA}.
It paves the path for a
better understanding of these positron factories in the Galaxy, and
their potential contribution to the local positron excess above 10
GeV as well as to Ultra-High-Energy cosmic rays.  The hard radiation
component is also a new tool for probing the role of magnetic
reconnection as an acceleration process in isolated pulsars, with
possible implications for other highly magnetised plasma in diverse
astrophysical contexts., e.g., black hole magnetospheres and
jet-accretion disc systems.

\bibliography{Vela20TeV_NatureA_V1_arxiv}

\bigskip

\section*{Acknowledgments}
The support of the Namibian authorities and of the University of
Namibia in facilitating the construction and operation of H.E.S.S.
is gratefully acknowledged, as is the support by the German
Ministry for Education and Research (BMBF), the Max Planck Society,
the German Research Foundation (DFG), the Helmholtz Association,
the Alexander von Humboldt Foundation, the French Ministry of
Higher Education, Research and Innovation, the Centre National de
la Recherche Scientifique (CNRS/IN2P3 and CNRS/INSU), the
Commissariat à l’énergie atomique et aux énergies alternatives
(CEA), the U.K. Science and Technology Facilities Council (STFC),
the Irish Research Council (IRC) and the Science Foundation Ireland
(SFI), the Knut and Alice Wallenberg Foundation, the Polish
Ministry of Education and Science, agreement no. 2021/WK/06, the
South African Department of Science and Technology and National
Research Foundation, the National Commission on Research, Science \& Technology of Namibia (NCRST),
the Austrian Federal Ministry of Education, Science and Research
and the Austrian Science Fund (FWF), the Australian Research
Council (ARC), the Japan Society for the Promotion of Science, the
University of Amsterdam and the Science Committee of Armenia grant
21AG-1C085. Work at NRL is supported by NASA. 

We appreciate the excellent work of the technical
support staff in Berlin, Zeuthen, Heidelberg, Palaiseau, Paris,
Saclay, Tübingen and in Namibia in the construction and operation
of the equipment. This work benefited from services provided by the
H.E.S.S. Virtual Organization, supported by the national resource
providers of the EGI Federation.
This research made use of the Python packages Astropy~\cite{astropy} and
naima~\cite{naima}. 
A.~Djannati-Ata\"i thanks Benoît Cerutti and Jérôme Pétri for fruitful discussions
on the striped-wind model during the
``Entretiens sur l'observation et la modélisation des pulsars'' sessions funded by
the Programme National Hautes Energies (PNHE) of which the support is acknowlegded here.
The authors wish to acknowledge the seminal role played by our late colleague, Okkie de Jager,
in opening up the VHE pulsar window of research.

\section*{Author Contributions Statement}
A.~Djannati-Ata\"i led the \hess project of the Vela pulsar and the main \hess data analysis.
G.~Giavitto and L.~Mohrmann performed the cross-check analyses used in this study along with
V.~Marandon. The statistical assessment of the results relies on Monte Carlo simulations perfromed
by M.~Spir-Jacob. 
A.~Djannati-Ata\"i developed the interpretation and modelling together with B.~Rudak,  M.~Spir-Jacob,
T.~Tavernier, E.~de~Ona~Wilhelmi and C.~Venter. 
The manuscript was prepared by A.~Djannati-Ata\"i, B.~Rudak,  E.~de~Ona~Wilhelmi, C.~Venter, and L.~Mohrmann.
T.~Lohse and M.~B\"ottcher supervised the review and discussion
of the manuscript among the coauthors.  The sketch in Fig.~4 was designed by H.~Prokoph together
with E.~de~Ona~Wilhelmi, based on an initial proposal by M.~Spir-Jacob and A.~Djannati-Ata\"i.
The other coauthors contributed by preparing and obtaining the
observations, calibrating the data, simulating showers and developing
analyses, developing, constructing, operating, and maintaining telescopes,
cameras, and calibration devices, conducting data handling, data reduction
and data analysis software.
The ephemeris used for phase-folding the \hess data was provided by M. Kerr, S. Johnston, R.M. Shannon and D. Smith. 
All authors meet the journal’s authorship criteria and have reviewed, discussed,
and commented on the results and the manuscript.

\section*{Ethics declarations}
The authors declare that they have no competing interests.

\section*{Data and Materials Availability}
The \hess{} raw data and the code used in this study are not public but belong to the H.E.S.S. collaboration. 
The high level data for the light curve (cf. Fig.~1), 
and the confidence interval for the spectral energy distributions (cf. Fig.~3) are available  at: \url{https://www.mpi-hd.mpg.de/hfm/HESS/pages/publications/auxiliary/2023_Vela_MultiTeV}.

\noindent


\appendix
\baselineskip24pt

\renewcommand{\thetable}{M\arabic{table}}
\renewcommand{\thefigure}{M\arabic{figure}}
\renewcommand{\theequation}{M\arabic{equation}}
\renewcommand{\theHfigure}{A\arabic{figure}}

\setcounter{figure}{0}

\newpage
\section*{Methods}
\subsection*{\hess{} Observations and Data Analysis}

Observations of the Vela pulsar were performed with the \hess{}  array of imaging Cherenkov telescopes, located in 
the Khomas Highland of Namibia   ($23^{\circ}16\arcmin18\arcsec$\,S, $16^{\circ} 30\arcmin 00\arcsec$\,E, 1800\,m).
The \hess{} array has been designed for the detection of high-energy (HE) and very-high-energy (VHE) \grs{} in the 10\,GeV-100\,TeV range. 
It consists of four imaging atmospheric Cherenkov telescopes (CT1-4), each having a $108~\rm m^2$
mirror area, placed in a square formation with a side length of
120 m, and a fifth telescope (CT5) with a larger mirror area of $614\, \rm m^2$ placed at the center. 
The latter telescope was added in 2012 to extend the energy range of the array to below 100\,GeV. 
The observations used for this study focused on the highest energy events and were performed in stereoscopic mode with CT1-4. 
Our first set of \gar{} observations of the Vela pulsar with CT1-4  
consisted of 16.3 hours and resulted in upper limits above a threshold energy of 170~GeV\cite{HessPSRULs2007}.

A total of 80 hours of data from 2004 to 2016 observing seasons were selected based on weather conditions and the instrumental status.
Observations were mostly performed in wobble mode \cite{hess:Crab} with a source-to-center distance of $0.7^\circ$, and with the zenith angle ranging between 20$^\circ$ and 40$^\circ$. When penetrating the atmosphere, \grs{} as well as charged cosmic rays interact with its constituents, 
producing showers of ultra-relativistic particles that emit Cherenkov light along their path in the air.
The light collected by each dish forms an image of this shower and is recorded by highly sensitive cameras consisting of photo-tubes and fast electronics.   
The data analysis starts with the reconstruction of the direction and the virtual impact point on the ground of each event, derived from the 
combination of information in shower images recorded by the camera of each telescope \cite{Hofmann99,hess:Crab}. 
The energy of each event and the discrimination parameters used to reject the background of charged cosmic rays that remain after a spatial (angular) cut 
at the 68\% containment radius of the instrument, 
are obtained via a multi-variate analysis \cite{apcmva2011} based on a boosted decision tree (BDT) classifier implemented within the \textsc{TMVA} package \cite{TMVA}.
The BDT is trained using extensive Monte Carlo (MC) simulations of $\gamma$-ray induced images \cite{SimTelArray} and
real {\it off}-source data as signal and background inputs, respectively.   
The results presented in this paper were cross-checked with an alternative calibration, and with two additional analysis chains for the reconstruction and background suppression 
\cite{DenauroisRolland2009, Ohm2009, Parsons2014}.

\subsubsection*{Timing and Phase Selection} 
The arrival time of each event is provided by a GPS receiver in the central trigger system of \hess{} and is then software-corrected
for the time delays in the array.  A long-term stability of better than 2 $\mu$s is achieved for the system \cite{TriggerFunk2004}. 
The pulsar phase corresponding to the arrival time of each event is calculated using the \textsc{Tempo2} package \cite{TempoGeneral}.  
Event arrival times provided by a GPS receiver in the central trigger system of \hess{} are transformed to the solar system barycentre 
where the pulsar phase of each event is computed using an ephemeris derived from 
radio data from the Parkes Radio Telescope. The ephemeris consisted of two overlapping solutions, valid for the ranges \texttt{MJD}~51602.43-56555.73 
and 54175.52-57624.20 (with fiducial phase references, \texttt{TZRMJD} = 54091.726 and 55896.55), 
with a precision of a few  milli-periods ($100-300 \,\mu $s). Vela is known for its recurrent glitches.
The two timing solutions are phase-connected and take properly into account the 3 glitches recorded in the years from 2004 to 2013 
at \texttt{MJD}~53193, 53960 and 55408.8. None of these glitches took place during an observation run and
the glitch at \texttt{MJD}~57734.5 (December 12, 2016, studied in detail in \cite{vela2016glitch})
lies beyond the  \hess 2016 observing period which ended at \texttt{MJD}~57541.7 (June 2, 2016).

\subsubsection*{Periodicity Search}
The search for pulsations was conducted 
at four predefined and increasing energy thresholds of 0.5, 1, 3 and 7 TeV. These energies were
intended to cover the plausible range of the source spectrum hardness (from soft to hard),
given the absence of an {\it a priori} knowledge of it.
Three types of periodicity tests were used: the H-test \cite{deJager89} where no \textit{a priori}
knowledge of the light curve (or phasogram, i.e., the phase-folded distribution of
events) is assumed, the C-test \cite{deJager1994} where the position and the
(approximate) width of the pulse shape are supposed to be known beforehand, and 
a maximum likelihood-ratio (LR) test \cite{LiAndMa83} based on \textit{a priori} defined
\textit{On}-- and \textit{Off}-phase intervals. 
The pulse P2 of the Vela pulsar, dominating in the
tens of GeV energy range, was considered and its parameters were
derived from the \fermi-LAT phasogram above 10 GeV \cite{VelaMonoPaper}, i.e., 
$\phi_{\rm P2}^{\rm GeV}=0.565$, and $w= 0.025$ (FWHM). The \fermi-LAT phasogram above 10 GeV was also
used to define the {\it On}- and {\it Off}-phase intervals as [0.55,0.6] and [0.7,1.0], respectively.

The pulsed nature of the signal enables one to extract the \textit{On}- and \textit{Off}-source events from the same
portion of the field of view, thereby eliminating one of the main sources of systematic effects arising from variations of 
acceptance as a function of direction in the sky and/or position in the camera. The required minimal significance
level for detection 
of pulsations is consequently defined as $4\sigma$, i.e., lower than that usually used for a DC (i.e. unpulsed) signal ($5\sigma$). 
Given the small number of events, the probability distribution function of all tests was computed numerically using extensive MC simulations~\cite{Jacob2019}.

The periodicity test results are given in Table~\ref{tab::HessVelaStatistics}.
The C-test resulted in trials-corrected (pre-trials) significance levels
of 4.9$\sigma$ (5.4$\sigma$) and  5.6$\sigma$ (6.0$\sigma$) 
above energy thresholds of 3 and 7 TeV, respectively.
For these thresholds, the corresponding H-test results are 3.9$\sigma$ (4.5$\sigma$) and  4.5$\sigma$ (5.0$\sigma$), while the likelihood-ratio test yielded post-trial significance levels of 4.7$\sigma$ and 4.8$\sigma$, with excess counts of 18.2 and 14.3 events, respectively.  
The total number of trials is conservatively assumed to be equal to 12 and corresponds to the number of 
tests for periodicity (i.e. 3: the C-test, H-test and the likelihood-ratio test) applied to the data
multiplied by the number of data sets (i.e. 4 sets corresponding to the 4 energy thresholds used for selection of events)\footnote{The number of trials is conservative for two reasons: (i) the 4 samples have overlapping energy ranges; and (ii) the 3 periodicity tests do not amount to 3 plain trials as they use exactly the same data sample. }.
Lower post-trial significance levels were obtained for the data sets with energy thresholds of 0.5 TeV and 1 TeV, e.g., for the C-test, 3.7$\sigma$ and 4.3$\sigma$, 
respectively, pointing to a hard energy spectrum at P2. 
Additional post-detection C-tests (likelihood-ratio tests) were performed above energy thresholds of 5 and 20 TeV. 
They resulted in significance levels of 5.8$\sigma$ (5$\sigma$, 18.2 excess counts) and 4.6$\sigma$ (4.3$\sigma$, 6.7 excess counts), respectively, confirming 
the hard photon spectrum.

\begin{table}[htb]
\centering              
\begin{tabular}{c | c  c  c |c }        
\hline                        
 Threshold Energy & C-test     & H-test     &  \multicolumn{2}{c}{LR}         \\    
 \hline
  (TeV)    & $\sigma$        &  $\sigma$        &   $\sigma$         & Excess     \\    
\hline                        
0.5         &  3.7 (4.3) & 2.7 (3.4)  &  3.8 (4.4)   & 23.2  \\      
1           &  4.3 (4.8) & 2.1 (3.0)  & 3.8 (4.4)    & 19.7 \\
3           &  4.9 (5.4) & 3.9 (4.5)  & 4.7 (5.2)    & 18.2 \\
5           &  5.8 	 & 4.7	      & 5.0          & 14.3 \\
7           &  5.6 (6.0) & 4.5 (5.0)  & 4.8 (5.3)    & 14.0 \\
20          &  4.6	 & 3.1	      & 4.3 	     & 6.7  \\
\hline                                   
\end{tabular}
\vspace{0.5cm}
\caption{Significance levels 
obtained for different periodicity tests: the C-test,
the H-test, and a maximum likelihood-ratio (LR) test.
The parameters for the C-test ($\phi_{\rm P2}^{\rm GeV}=0.565$, and $w= 0.025$ (FWHM)) and for
the LR test ({\it On}- and {\it Off}-phase intervals defined as [0.55,0.6] and [0.7,1.0], respectively)
were derived from the \fermi-LAT phasogram above 10 GeV \cite{VelaMonoPaper}. 
The tests were applied on data selected above increasing energy thresholds, four of which, 0.5, 1, 3 and 7 TeV,
were defined {\it a priori} to search for pulsations.
Their initial significance level (shown in parenthesis) is corrected for the number of trials which has been conservatively estimated to be 12 (4 energy thresholds and 3 tests, see text).
The number of events in excess of the background quoted in the last column was estimated with the LR test.}
\label{tab::HessVelaStatistics}
\end{table}

\subsubsection*{Light curve Fitting}
The characterization of P2 was performed via an unbinned likelihood fit of an asymmetric
Lorentzian function~\cite{abdo2010, VelaMonoPaper}.
The fit to data selected above 5~TeV resulted in a position 
$\phi_{\rm P2}^{\rm TeV}=0.568^{+0.003}_{-0.003}$, and a sharp outer edge (or trailing edge), $\sigma_{\rm T}^{\rm TeV}=0.004^{+0.006}_{-0.004}$, both of which are compatible 
with the fitted values obtained above 20~GeV, i.e., $\phi_{\rm P2}^{\rm GeV}=0.565\pm 0.001$ and $\sigma_{\rm T}^{\rm GeV}=0.003\pm0.001$ \cite{VelaMonoPaper}. 
The central fitted value of the inner edge width (or leading edge) of the TeV pulse, $\sigma_{\rm L}^{\rm TeV} =0.007^{+0.007}_{-0.004}$, was found to be slightly smaller than 
$\sigma_{\rm L}^{\rm GeV}=0.017\pm0.002$, but the difference is not statistically significant ($< 1.5 \sigma$).

The fact that only P2 is detected in the multi-TeV range is consistent with the evolution 
of the phasogram with increasing energy. Indeed,
the ratio of P1~/~P2 amplitudes
decreases with increasing energy, with P1 dominating below 300 MeV, while P3
 dims and slides to later phases with increasing energy \cite{abdo2010}. 
This trend was confirmed in the tens of GeV range \cite{VelaMonoPaper}, P2 being the sole significant feature in Vela's phasogram there.

\subsubsection*{Spectral Derivation}
Data were selected for the P2 and {\it Off}-phase intervals, defined as [0.55-0.6] and [0.7-1.0], respectively. 
The energy spectrum was derived using a maximum likelihood fit within a forward-folding scheme, assuming \textit{a priori} spectral models~\cite{Piron2001}.
Instrument response functions (IRFs) were computed through extensive MC simulations as a function of the energy, zenith and azimuthal angles of the telescope pointing direction, the impact parameter of showers, and the configuration of the telescope array for each observing period.      

The fit of a power law to the overall data set in the $660\,\mathrm{GeV} - 28.5\,\mathrm{TeV}$ energy range resulted in a very hard spectrum with photon
index $\Gamma_{\rm VHE} = 1.4 \pm 0.3^{\rm stat} \pm 0.1^{\rm syst}$ and normalization
$\Phi_0= \left(1.74 \pm 0.52^{\rm stat} \pm 0.35^{\rm syst}\right) \times 10^{-15}$ {erg}$^{-1}$cm$^{-2}$s$^{-1}$
at the decorrelation energy $E_0 = 4.24 $~TeV.
This corresponds to an isotropic luminosity $L_{20 \rm TeV} \simeq 2 \times 10^{30} {\rm erg \,s^{-1}} $ for the source distance of 287~pc\cite{Dodson2003}.
The systematic uncertainties on these results have been adopted from the study carried out in \cite{hess:Crab}.
The limited statistics do not allow a test for a statistically significant deviation from the power-law hypothesis. 
We adopt conservatively 20~TeV as the maximum detected energy for individual photons in the following sections,
noting that the energy spectrum extends up to 28.5~TeV due to events displaying an energy beyond 20~TeV.

\newpage
\subsection*{Supplementary Material}
\label{sec::ToyModels}

\subsubsection*{The Multi-TeV Component}
\label{sec::ic}
The TeV emission is most likely produced by inverse-Compton (IC) scattering of low-energy photons by energetic
electrons.  The target photons in Vela might consist of the observed
non-thermal X-rays~\cite{rxte2002,manzali2007}, UV~\cite{uv2005}, and
optical~\cite{velaoptical2007,velaoptical2019} to near-infrared
\cite{zyu2013} emission (O-NIR), or thermal X-rays from the neutron
star surface \cite{manzali2007}.
The spectral measurements from UV to NIR have shown a flat $F(\nu)$
spectrum ($\alpha_\nu=0.01$) in the range $\epsilon \in  [0.6, 10]\,$ eV
($\log_{10}(\nu/\rm{Hz} )=14.15-15.4$)\cite{zyu2013}.
These photons are generally considered as being
emitted through SR of secondary pairs. 
In the magnetospheric scheme, secondary pairs are produced along the outer gaps 
(e.g. \cite{romani96, RudakDyks2017}), or between the NS
surface and $\sim 0.5\, R_{\rm LC}$, as assumed in
\cite{ahvela2018}. In the wind-based framework, pairs are produced
around the current sheet and their synchrotron emission develops as an
isotropic radiation field in the optical to IR domains, as first
proposed in \cite{Lyubarsky1996} and investigated in recent PIC
simulations \cite{Philippov2018, Hakobyan2019}.

The scattering regime depends
on the target photon energy in the center-of-momentum frame, $\gamma
\epsilon$, and takes place in the Thomson or deep Klein-Nishina (K-N)
regimes for $\gamma \epsilon \ll m_{\rm e} c^2$ and $\gamma \epsilon
\gg m_{\rm e} c^2$, respectively. The lowest energy photons measured from
Vela lie in the NIR domain at an energy of $\sim0.6$ eV, together with two other measurements at 0.33 and 0.2 eV,
though with lower precision \cite{zyu2013}.
This means that for the lower bound of the (measured) NIR radiation field   
the scattering takes place already in the mildly relativistic case.
The luminosity in the O-NIR range, $L_{0.6\,\rm eV} \simeq \omega_{\rm IR} \,2.3 \times 10^{28}\, {\rm erg \,s^{-1}}$,
(where $\omega_{\rm IR} <1$ corrects for the solid angle) is two to three orders of magnitude 
below that in the thermal, $L_{\rm X}^{\rm th}\simeq 8 \times 10^{31}\, {\rm erg \,s^{-1}}$, and non-thermal X-rays, $L_{\rm X}\simeq \omega_{\rm X} \,10^{31}\, {\rm erg \,s^{-1}}$.
However, for photons beyond the optical range, e.g. with energy $\epsilon> 10$~eV,  the IC emissivity is strongly suppressed due to the severe drop 
in the IC scattering cross section,  $\sigma_{\rm KN}/\sigma_{\rm T} \lesssim 10^{-5}-~10^{-6}$. 
Hence the O-NIR photons are the dominating targets for efficient IC emissivity, noting
that some contribution is also expected from the extreme K-N regime \cite{RudakDyks2017}.
The VHE photon energy  $E_{\rm VHE} \simeq \gamma\, m_{\rm e}\, c^2$ is thus set by $\gamma$.

The highest photon energies observed by \hess{} imply a lower
limit on the maximum particle energies of $\gamma_{\rm IC}^{\rm max} \gtrsim 4 \times 10^7
\, (E_{\rm VHE}/20\,{\rm TeV})$.
When taking into account the spectral shapes of both GeV and TeV components, 
Lorentz factors of $\gtrsim 7\times 10^7$ are needed to reproduce the data 
in the TeV range (see below).

\subsubsection*{The GeV Component}
When considering pulsar gaps, 
   electrons are believed to be accelerated by
  the component of the electric field parallel to the local magnetic field, $E_{||}$,
  and radiating 
  in the GeV regime mainly through CR.
   The maximum achievable Lorentz factor $\gamma^{\rm max}_{\rm CR}$ is limited by radiation reaction
   and not by escape from the acceleration region.
  Equating the energy loss and gain rates
($\frac{2}{3} \,\frac{e^2 c}{\rho_{\rm c}^2}\,\gamma^4$ = $e\,c\,
E_{||}$), results in an expression of $\gamma_{\rm c}^{\rm CR}$ as a
function of the magnetic field $B$ and curvature radius $\rho_{\rm c}$,
as $\gamma_{\rm CR}^{\rm max}$ = $\left(\frac{3\,\eta
  B}{2\,e}\right)^{1/4} \rho_{\rm c}^{1/2}$, where $\eta$ is the magnetic conversion efficiency, 
$E_{||} = \eta \,B$ \footnote{For simplicity 
we use a constant value for $E_{||}$ along the field lines in the gap, as usually done in magnetospheric models.}.
The  magnitude of $\eta$ depends on the particular version of the acceleration
gaps 
with values usually assumed to be below 10\% \cite{Hirotani2006,ahvela2018} at the LC, although 
in some models values as large as 30\% are used \cite{hirotani2013,Takata2006}.
The radius of curvature can
be expressed in units of the LC radius ($\rho_{\rm c}=\xi\, R_{\rm
  LC}$, with $R_{\rm LC}=cP/2\pi$, and $P$ the period of the
pulsar). Assuming a static dipole geometry for the magnetic field,
$B(r)=B_{\rm ns}(R_{\rm ns}/r)^3$, with $B_{\rm ns}$ and $R_{\rm ns}$
the surface magnetic field and the neutron star radius, respectively,
$\gamma_{\rm CR}^{\rm max}$ can be expressed as $\gamma_{\rm
  CR}^{\rm max}=({3 \pi/c\,e})^{1/4} \xi^{1/2}\eta^{1/4} B_{\rm ns}^{1/4}R_{\rm
  ns}^{3/4} P^{-1/4}$.

Using the inferred value of $B_{\rm ns}$, the assumed value of $R_{\rm ns}$ and the measured period  $P$ for Vela
($2.47\times10^{12}$~G, 12~km and 89~ms, respectively),  $\gamma_{\rm CR}^{\rm max}$  
and the corresponding CR photon energy $E^{\rm max}_{\rm CR}$ for emission regions close to
the LC can be written as:

\begin{equation}
\label{Eq::cr_gamma_max_1}
\gamma_{\rm CR}^{\rm max} \simeq 4 \times10^7 \, \xi^{1/2}\eta^{1/4}_{-1}\\
\end{equation}  
\begin{equation}
\label{Eq::cr_gamma_max_2}
E^{\rm max}_{\rm CR}\simeq 5  \,\rm{GeV} \,\xi^{1/2}\eta^{3/4}_{-1}\\ 
\end{equation}  
\noindent with $\eta_{-1}= \eta/0.1. $

Assuming that CR is the dominating radiation process forming the HE
spectral component, a mono-energetic beam of particles with  $\gamma =
\gamma^{\rm max}_{\rm CR}\sim 4\times 10^7$  would contribute near 
E$^{\rm  peak}_{\rm CR}\sim E^{\rm max}_{\rm CR}$.
Given the measured  E$^{\rm  peak}_{\rm HE}\simeq$1.5\,GeV \cite{fermifirstvela,abdo2010,VelaMonoPaper}, 
for an emission taking place near the LC, $\xi \sim 1$, values of $\eta < 0.1$ and 
Lorentz factors $<4\times 10^7$ are implied (see Fig.~\ref{fig::etaxigamma}). 
An estimate of the number of the contributing electrons can be obtained from the
inferred luminosity $L_{1.5 \,\rm GeV} \simeq \omega_{\rm HE}\, 9 \times 10^{33} \,{\rm erg \,s^{-1}}$
(where $\omega_{\rm HE} < 1$ corrects for the solid angle) and given
  the curvature energy loss rate, 
  $\frac{-dE}{dt}\rfloor_{\rm CR} = \frac{2}{3} \,\frac{e^2 c}{\xi^2  R_{\rm LC}^2} \,\gamma^4 \simeq 6.5\times 10^4
  \,(\gamma/(4\times 10^7))^4 \, \xi^{-2} \,{\rm erg \,s^{-1}}$
 as:
$N_0^{\rm CR} \sim L_{1.5 \rm GeV}/(\frac{-dE}{dt}\rfloor_{\rm CR}) \simeq\ \omega_{\rm HE} \,
  \xi^{2} \,1.4 \times 10^{29} \,{\rm particles} $.  

Alternatively, the GeV component can be interpreted as SR, if
considering non-ideal MHD plasma conditions ($E>B_\bot$) deep in the reconnection layers to promote
the maximum energy beyond the maximum reachable synchrotron energy ($\simeq$160
MeV)\cite{kirk2004}. In such a scenario, the peak of the GeV component would correspond
to the characteristic energy $E_{\rm SR}^{\rm max} \sim E_{\rm HE}^{\rm
  peak}\simeq 1.5\,$GeV, with
\begin{equation}
\label{eq::SR}  
\gamma^{\rm max}_{\rm SR} \simeq 1.3 \times 10^6 (B_\bot/B_{\rm LC})^{-1/2}\,
(E_{\rm SR}^{\rm max}/ 1.5\,{\rm GeV})^{1/2}
\end{equation}  
where $B_{\rm LC}= 5.53\times 10^4 {\rm G}$ is the magnetic field
intensity at the LC.

The SR energy loss rate, $\frac{-dE}{dt}\rfloor_{\rm SR}) \simeq 5\times
10^6\,(\gamma/1.3 \times 10^6)^2\,{\rm erg \,s^{-1}}$ is two orders of magnitude larger than in the CR case,
which implies in this case a smaller number of particles,  $N_0^{\rm SR} \sim  L_{1.5 \rm
  GeV}/(\frac{-dE}{dt}\rfloor_{\rm SR}) \simeq \ \omega_{\rm HE} \, 1.8 \times 10^{27}$, involved in the GeV
radiation.

\subsubsection*{The CR/IC and SR/IC scenarios}
\label{sec::luminosity}
Given the phase alignment of the GeV and TeV pulsations, we assume 
that the same population of electrons, within an energy range partially or completely in overlap, 
and from similar spatial regions, is at the origin of the two components.
The emission regions do not necessarily have to be identical as photons  
from radially extended zones can arrive at Earth at similar phases, i.e. 
form caustics, thanks to special-relativistic effects and the $B$-field structure, 
within (or slightly beyond) the magnetosphere  \cite{morini1983,RY1995,crz2000,dr2003,dhr2004,BaiSpitkovsky2010a},
or within the equatorial CS  in the near wind zone \cite{cps2016,Philippov2018}.

Using the results obtained above,
we can derive some constraints on the target radiation field energy density, and thereby,
on the effective volume of the IC emission region  $V_{\rm IC}$ in both
SR/IC  and CR/IC scenarios. 
We first consider IR to optical target photons 
and restrict the computations to the deep K-N regime, i.e. $\gamma \, \epsilon \sim 10 \times m_e\,
c^2$, and hence the energy range
to $(\epsilon_{\rm min},\epsilon_{\rm
  max})=(0.1,4)$~eV.  
The energy loss rate in the extreme K-N limit is given by~\cite{1970RvMP...42..237B}:
$\frac{-dE}{dt}\rfloor_{\rm IC}\simeq \zeta \, (3/8)\, \sigma_{\rm T}\,c\,
  m_e^2\,c^4 \int_{\epsilon_{\rm min}}^{\epsilon_{\rm max}}
  \frac{n({\epsilon})}{\epsilon} [\log(\Gamma_e)-11/6] \,d\epsilon $
where $\zeta < 1$ represents an overall efficiency factor for the IC scattering
(i.e. anisotropy, target photon direction/opening angle with respect to the
accelerating field direction, etc),
$\Gamma_e=\frac{4\, \epsilon \, \gamma}{m_{\rm e} c^2} $, and $n(\epsilon)$ is the 
target photon density~\footnote{The spectral fits use a more detailed calculation including an extension to $\epsilon_{\rm min}=0.005$~eV and 
using the full K-N cross section formula.}.

The target radiation field displays a photon index of
$\alpha=\alpha_\nu+1=1.01\pm0.01$ in the optical to IR range, 
and a luminosity of $L_{0.6\,\rm eV} \simeq \omega_{\rm IR} \,2.3 \times 10^{28}\,
{\rm erg \,s^{-1}}$ \cite{zyu2013}.  
The target photon density $n(\epsilon)$ depends inversely on the effective interaction volume $V_{\rm IC}$.
This is also the case for the IC luminosity  
$L_{IC}^{\rm KN}= N_0^{IC}\,\frac{-dE}{dt}\rfloor_{\rm IC} =  L_{20 \rm TeV} \simeq  \omega_{\rm VHE} \,2 \times10^{30} {\rm erg \,s^{-1}} $.
Assuming the same solid angle for the GeV and TeV emissions, $\omega_{\rm HE}=\omega_{\rm VHE}$, and using 
the number of emitting particles $N_0^{\rm CR}$ and $N_0^{\rm SR}$ derived above in the CR and SR scenarios,   
the required energy density of the target field could be expressed for each case as:
 
\begin{equation}
\label{Eq::Ueps_ic_KN_cr}
U_{\epsilon}^{\rm CR} \simeq 3.8\times10^{13}\,{\rm eV} \,{\rm cm^{-3}} \frac{1}{\zeta \,\xi^2}\, \left(\frac{L_{IC}^{\rm KN}}{ 2\,10^{30} 
\,{\rm erg/s}}\right)\left(\frac{N_0^{\rm CR}}{1.4\,10^{29}}\right)^{-1}
\end{equation}
and
\begin{equation}
\label{Eq::Ueps_ic_KN_sr}
U_{\epsilon}^{\rm SR} \simeq 3.0\times10^{15}\,{\rm eV} \,{\rm cm^{-3}}   \frac{1}{\zeta}\, \left(\frac{L_{IC}^{\rm KN}}{ 2\,10^{30} 
\,{\rm erg/s}}\right)\left(\frac{N_0^{\rm SR}}{1.8\,10^{27}}\right)^{-1}
\end{equation}

Assuming a crossing time $t_{\rm c}= \tau \, R_{\rm LC}/c $, where $\tau <1$ marks the uncertainty 
on emitting zone dimensions, and writing  $U_{\epsilon}=L_{0.6\,\rm eV}\times t_{\rm c} / V_{\rm IC}$,
one obtains the following constraints:

\begin{equation}
\label{Eq::V_ic_KN_cr}
V^{\rm CR}_{\rm IC}\sim  \upsilon \, \tau \, \xi^2 \,\zeta\, \omega_{\rm IR}\, 5.3 \times 10^{24}{\rm cm}^3
\end{equation}
\begin{equation}
\label{Eq::V_ic_KN_sr}
V^{\rm SR}_{\rm IC}\sim \upsilon \, \tau \, \zeta\, \omega_{\rm IR} \, 6.8 \times 10^{22}{\rm cm}^3
\end{equation}

\noindent where $\upsilon > 1 $ is a correction factor to take into account 
the impact of the lower bound of the target photon
energy range on the IC luminosity, e.g. a factor as high as $\upsilon \sim 10$ for 
$\epsilon_{\rm min}=0.005$~eV, as illustrated through spectral fits further below.
Considering a particle density  $n_{\rm tot}^{\rm LC} \sim \kappa_4 \,n^{\rm LC}_{\rm GJ} \simeq \kappa_4 \,4.3\times 10^8 \,{\rm cm^{-3}}$ near the LC, 
where $n^{\rm LC}_{\rm GJ}$ is the Goldreich--Julian particle density and $\kappa_4=\kappa/10^4$ the pair multiplicity,
the effective volumes $V^{\rm CR}_{\rm IC}$ and $V^{\rm SR}_{\rm IC}$ contain the total number of particles
$N^{\rm CR}_{\rm tot} \sim n^{\rm tot}_{\rm LC} \, V^{\rm CR}_{\rm IC}\sim 2.3\times 10^{33}$ and   
$N^{\rm SR}_{\rm tot} \sim n_{\rm tot} \, V^{\rm SR}_{\rm IC}\sim 2.9\times 10^{31}$
for the CR/IC and the SR/IC scenarios, respectively. The particles contributing to the peak of 
the HE and VHE components represent then a fraction  $N_0^{\rm CR}/N^{\rm CR}_{\rm tot} \sim N_0^{\rm SR}/N^{\rm SR}_{\rm tot} \sim 10^{-4}$ (up to 
the respective correction factors in Eqs.~\ref{Eq::V_ic_KN_cr}, \ref{Eq::V_ic_KN_sr}) of the total
number of particles available near the LC.  

The effective volume under the CR/IC  hypothesis $V^{\rm CR}_{\rm IC}$ is of the same order as the  
SR emitting volume mentioned in \cite{ahvela2018}, i.e. 
$V^{\rm SR}\sim \frac{\pi\,(0.5 R_{\rm LC})^4}{2\,R_{\rm LC}}\,(r_{\rm out}^2-r_{\rm in}^2)\simeq 3 \times 10^{23}{\rm cm^3}$,
but the unknown magnitude of $\upsilon \, \tau \, \xi^2 \,\zeta\,\omega_{\rm IR}$ makes any further comparison difficult.   
In the SR/IC framework, target photons are produced around the current sheet
and the IC interaction can be assumed to be isotropic, hence $\zeta\, \omega_{\rm IR} \sim 1$.   
As suggested by PIC simulations, 
the layer thickness $\delta$ can be estimated 
from the fiducial Larmor radius of the electrons 
accelerated in the open field line region beyond the LC \cite{cerutti2017,Philippov2018},
$ \rho_L=\gamma\,m_{\rm e} c^2/(e\,B_{\rm LC})\sim \delta \simeq  4\times 10^4 \,{\rm cm}$
for $\gamma_{\rm c}^{\rm SR}\simeq 1.3 \times 10^6$.
This corresponds to a SR emission volume
$V^{\rm SR}\sim \delta \times R_{\rm LC}^2 \simeq 1.2\times 10^{22} {\rm cm}^3$, which is 
of the same order as the constraint in Eq. (\ref{Eq::V_ic_KN_sr}). 

\subsubsection*{Heuristic spectral models}
\label{sec::toy_model_results}

To further explore  the implications of the \hess{} data, we perform a joint fit 
to the HE and VHE components thus taking their spectral features into account.  
The limited statistics of the \hess{} measurement only allows the fitting of a power-law function to the data.
Consistently, we assume for the energy distribution of the IC emitting particles the functional form:
\begin{equation}
\label{Eq::ParticleDistribution}
 \frac{d^2N}{d\gamma dt} \propto (\gamma/\gamma_0)^{-p}\,{\rm exp}\left[-(\gamma/\gamma^{\rm max})^\beta\right]
\end{equation}

\noindent where the argument of the (super-) exponential cutoff represents 
$\gamma^{\rm max}_{\rm CR}$ or $\gamma^{\rm max}_{\rm SR}$, which are identified with $\gamma_{\rm IC}^{\rm max}$ 
in the CR/IC and SR/IC schemes, respectively.

\begin{figure}
\centering \includegraphics[width=12cm]{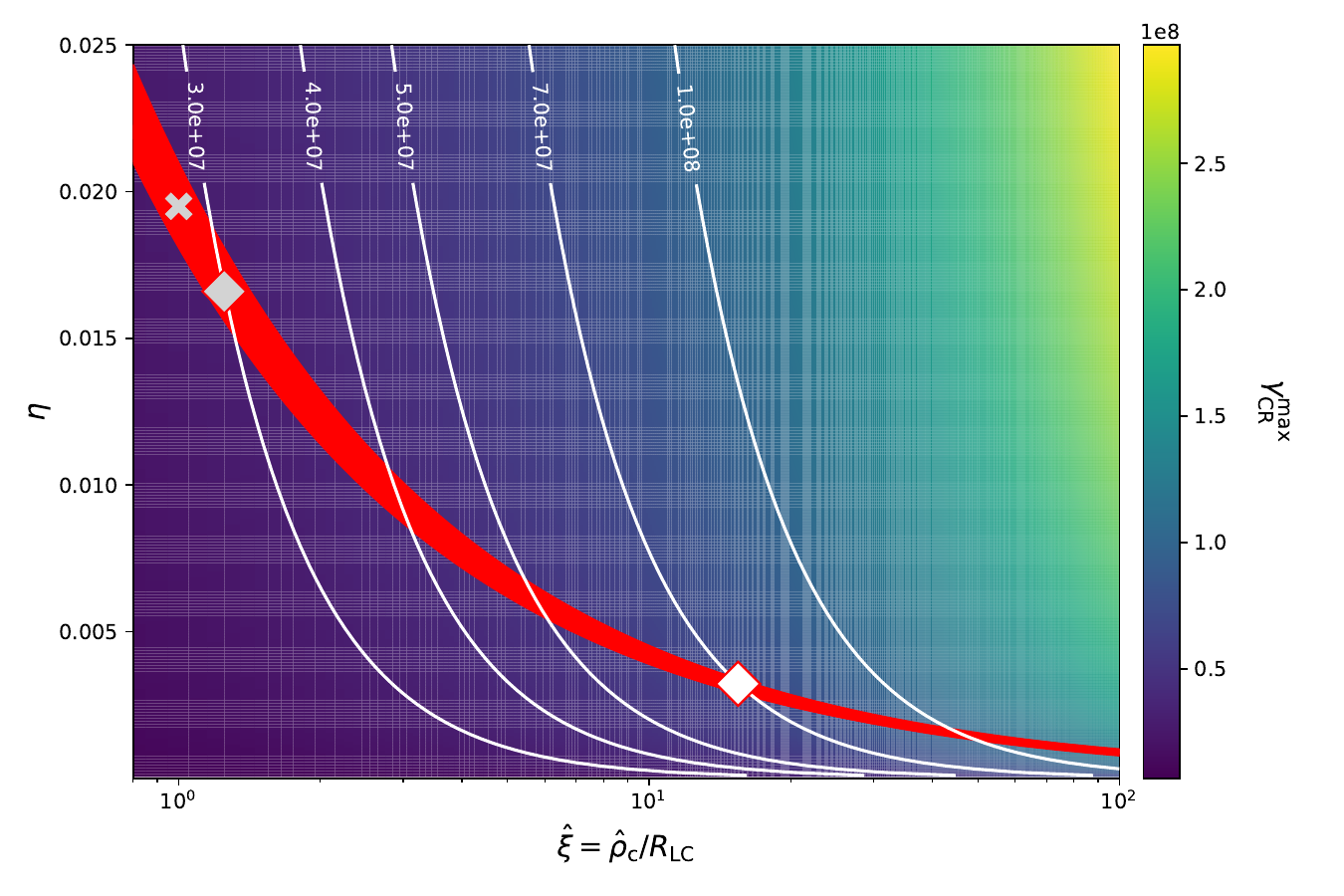}
\caption{ {\bf Constraints in the CR/IC scenario on magnetic conversion efficiency.}
The  maximum achievable Lorentz factor for electrons, $\gamma^{\rm max}_{\rm CR}$  is shown in color scale
(increasing contours levels are shown as white curves) as a function of
 magnetic conversion efficiency ${\eta}$, and scaled curvature radius, $\hat{\xi}$.
 The red curve shows $({\eta}, \hat{\xi})$ values when fitting the HE component, i.e. they correspond to particle trajectories which contribute
 most to the HE peak, $E^{\rm max}_{\rm CR}\sim E_{\rm HE}^{\rm peak} \simeq 1.5\,$GeV. The thickness of the red curve 
 illustrates the uncertainty on HE peak energy ($\sim \pm 10\%$). The gray cross shows the solution  $({\eta}, \hat{\xi})\sim(0.02, 1)$
 yielding $\gamma^{\rm max}_{\rm CR}\simeq 2.8 \times 10^7$, while the gray and white diamonds 
correspond to curves $\rm{I_a}$ and  $\rm{I_c}$ in Fig.~\ref{fig::spectral_models} of the main text with $\gamma^{\rm max}_{\rm CR} \simeq 3\times 10^7$ and $7\times 10^7$, respectively. } 
\label{fig::etaxigamma}
\end{figure}

\begin{figure}
\centering \includegraphics[width=12cm]{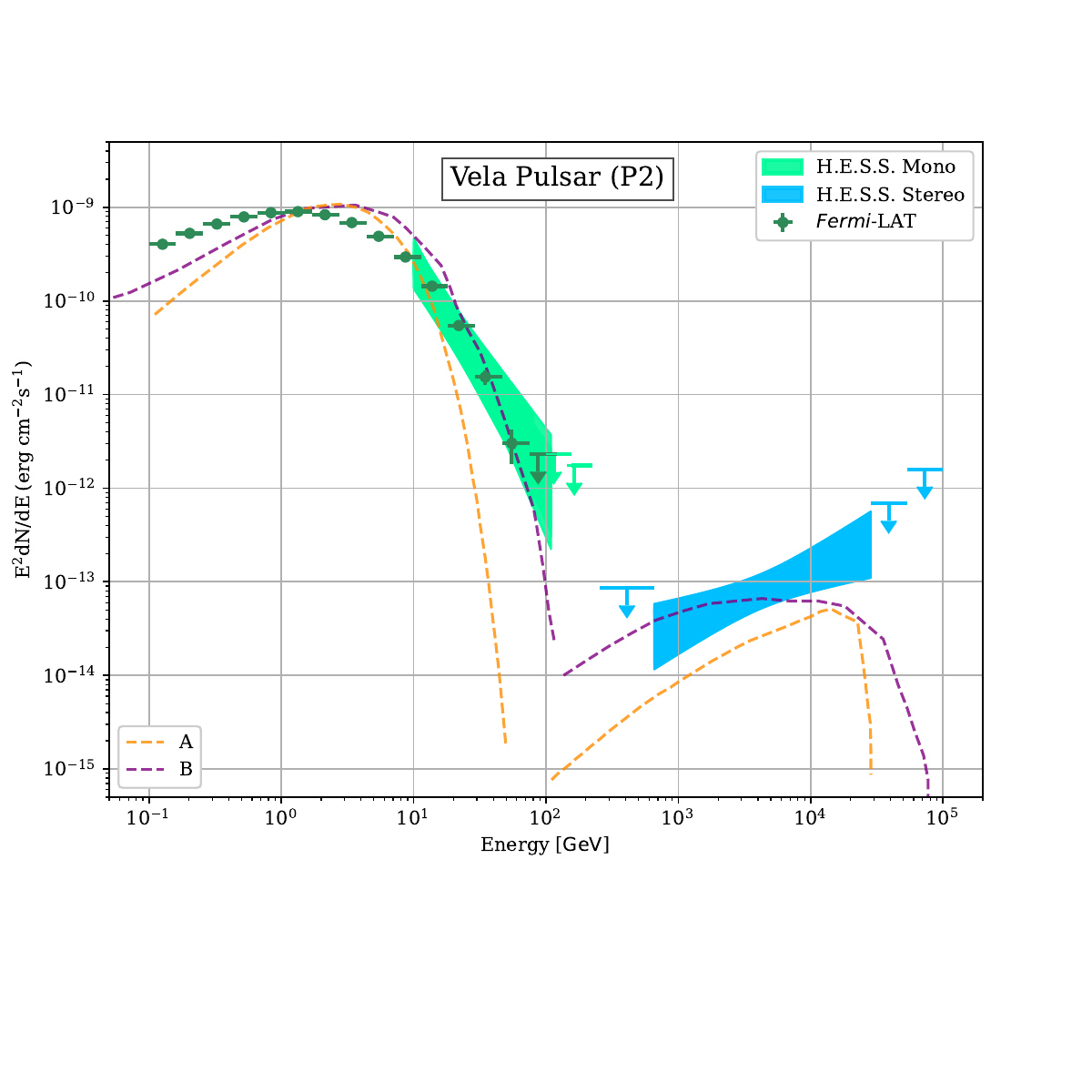}
\caption{ {\bf SEDs from two publications in the CR/IC and SCR/IC schemes compared to the Vela data.}
  {\it Data}: see Fig~\ref{fig::spectral_models}.
  {\it Curves}:
  Two recently published models
    are compared to the Vela SED.
  Curve A shows the phase-resolved SED
  of an outer-gap-based CR/IC model applied to Vela P2 \cite{RudakDyks2017}. 
  The curve labeled B corresponds to the re-scaled version of a phase-averaged computation based on
  a  separatrix/CS scenario \cite{ahvela2018} where  primary
  particles are predominantly accelerated in the CS, though not 
  through magnetic reconnection. Here, particles are cooled via SCR in the MeV and GeV ranges.   
  Both models predict a GeV peak energy ($E_{\rm HE}^{\rm peak} \gtrsim 3\,$GeV) larger than that measured by \fermi-LAT , i.e. they do not lie on
    the red curve in Fig.~\ref{fig::etaxigamma}, hence 
  providing for larger maximum Lorentz factors $\gamma^{\rm max}$
  for IC-emitting particles. 
  The target SR photon spectrum is extrapolated for both models to the FIR ($\epsilon_{\rm
    min}=0.001$ and $0.005$~eV, for curves A and B, respectively). }
\label{fig::sedmodels}
\end{figure}

In the CR/IC scenario, the inverse-squared dependence of the CR energy loss rate on radius of 
curvature, $-\frac{dE}{dt}\rfloor_{\rm CR} \propto {\rho_{\rm c}^{-2}} \propto \xi^{-2}$ (see above), 
implies a narrow distribution for the trajectories of particles contributing most to the GeV component, and hence also 
for the energy distribution of particles (see, e.g., Fig.~11 in \cite{Hirotani2006}).  
For the computation of CR,  we consequently limit the extent of the particle distribution at its lower energy bound to $\gamma^{\rm max}_{\rm CR}/10$, 
and define an effective radius of curvature $\hat{\rho}_c$ (or the scaled radius $\hat{\xi}$) representing the
particle trajectories which contribute to the peak near  $E_{\rm HE}^{\rm peak} \simeq 1.5\,$GeV, i.e. for which
$E^{\rm max}_{\rm CR} \sim 1.5\,$GeV. 
As discussed above, 
for a given $E^{\rm max}_{\rm CR}$ there is a degeneracy between values 
of $\eta$ and $\xi$, which also determine the maximum Lorentz factor $\gamma^{\rm max}_{\rm CR}$ (Eq.~\ref{Eq::cr_gamma_max_1}).  
We consider two cases. First, the emission is hypothesized to take place near the LC, i.e. $\hat{\xi}\sim 1$.  
The fit is constrained by the GeV data in this case and results in $\eta\simeq 0.02$ and $\gamma^{\rm max}_{\rm CR}\simeq 2.8 \times 10^7$
(shown by a gray cross in Fig.~\ref{fig::etaxigamma}).
The latter value, when identified to  $\gamma^{\rm max}_{\rm IC}$, is insufficient to reproduce 
the VHE data (see curve labeled $\rm{I_a}$ 
in Fig.~\ref{fig::spectral_models} of the main text).  
In the second case, the maximum Lorentz factor of IC-emitting particles is also constrained 
through the fit to the TeV component which results in $\gamma^{\rm max}_{\rm IC}\gtrsim 7\times 10^7$. 
Different combinations of ($\eta$, $\hat{\xi}$) can satisfy this condition, granted that $\eta\ll0.1$ and $\hat{\xi}\gg1$. These combinations
(which lie on the red curve in Fig.~\ref{fig::etaxigamma} to the right side of the white diamond)
imply a dissipation region located beyond the LC\footnote{Given the inclination and viewing 
angles in Vela ($\sim 70^{\circ}$ and $\sim65^{\circ}$, respectively), the open field lines of interest cross the LC surface at 
radial distances not much larger than $R_{\rm LC}$.}
where open field lines reach curvature radii well above $R_{\rm LC}$ 
and can  provide for acceleration to higher energies than traditionally assumed 
in the magnetospheric picture\footnote{We note, however, that more complicated schemes such as 
the non-stationary gap model as adopted in \cite{Takata2016} and having recourse to a combination of different particle injection rates could still accommodate the data.}.     
The solution corresponding to ($\eta\simeq 0.003$, $\hat{\xi}\simeq 15$) is marked as a white diamond in Fig.~\ref{fig::etaxigamma}
and is shown in Fig.~\ref{fig::spectral_models} of the main text in two versions: for the curve labeled $\rm{I_b}$ the IC component
is calculated using O-NIR target photon field, while for  $\rm{I_c}$ it is extended to the FIR. The comparison of
these two curves illustrates the impact of the target energy range on the TeV luminosity, 
i.e. a gain in luminosity of almost an order of magnitude for the extrapolated target energy range.    
The luminosity of the IC component depends also strongly on the effective interaction volume which is 
a complex quantity to tightly constrain. 
The normalization of the models is hence (almost) a free parameter in the spectral fits. 
On the other hand, the parameters of the energy distribution of the parent population can be constrained by the joint fit,
though not unambiguously given both the correlation between the spectral index $p$ and the exponential power $\beta$, and the limited extent of
the distribution itself towards lower energies (see above). For the two solutions ($\eta\simeq0.02, \hat{\xi}\simeq 1$) and ($\eta\simeq0.003,\hat{\xi}\simeq 15$) 
we obtain ($p,\beta) \simeq (0.6, 1.9) $
and  $\simeq(1.1, 2.0 )$, respectively. For both solutions, we note a power deficit in the lower energy part of the
HE spectrum ($< 1$~GeV) as compared to the data. This deficit is usually attributed to the SR contribution to this part of the spectrum (globally modeled as 
synchro-curvature radiation, SCR see e.g. \cite{2010ApJ...725.1903V}), which is not included in the spectral model.
Predicted spectral energy distributions (SEDs) taken from two recently published models
adopting more sophisticated CR/IC and SCR/IC schemes and including computation of
light curves \cite{RudakDyks2017,ahvela2018} are shown in Fig.~\ref{fig::sedmodels}\footnote{Inspired by the first preliminary announcement of a \hess{} multi-TeV signal \cite{adatexas2017}.}.

In the SR/IC scenario, the formation of a hard power-law particle spectrum, i.e. with index $p \in [1, 2]$,
is expected from PIC simulations of acceleration via
relativistic magnetic reconnection  (e.g., \cite{werner2016,werner2017,cps2016,Hakobyan2019}).
Accordingly, we use a larger energy span (few $10^5$) for particles as compared to the CR case.
As for the IC target photon field energy span, we limit the computations to the measured O-NIR domain.      
Using an exponentially cutoff power law for the particle energy distribution (see the functional form 
\ref{Eq::ParticleDistribution} given above), 
the fit of the resulting SR component to the GeV data yields shape parameters $p\simeq 1$, $\beta\simeq 1.8 $ and 
$\gamma^{\rm max}_{\rm SR} \simeq 1.3 \times10^6$ (see Eq.~\ref{eq::SR}). In this case, the spectral model fits the HE 
data in the $<1$~GeV range without requiring an additional SCR component.
The best-fit cutoff value with the (super-) exponential form is, 
however, well below the value $\gamma^{\rm max}_{\rm IC}\gtrsim 7\times 10^7$  required by the  \hess{} measurement (see curve labeled
 $\rm{II_a}$ in Fig.~\ref{fig::spectral_models} of the main text). 
For the SR/IC scenario to reproduce the \hess{} data, IC emitting particles must therefore have a significantly higher 
energy cutoff than the one deduced from the SR cooling. 
PIC simulations have confirmed that the acceleration and SR cooling processes take place 
in subsequent steps and different regions \cite{cerutti2013}. 
Recent studies including synchrotron cooling and pair production 
have further shown that the highest energy particles are not 
trapped by the magnetic loops (or plasmoids/magnetic islands)  but accumulate in their vicinity  where the magnetic field
intensity is weak and the SR cooling is inefficient \cite{Hakobyan2019}.
Hence they can reach energies well beyond the SR burn-off limit $\gamma^{\rm max}_{\rm SR}$ 
and exceeding the magnetization $\sigma_{\rm LC} \simeq 7\times 10^5 $ at LC\footnote{The  magnetization at the LC is defined
through Michel's magnetization parameter $\mu_{\rm M}$ as: 
{$\sigma_{\rm LC} \sim \mu_{\rm M} = {B_{\rm LC}^2}/({4\,\pi\,\kappa_{4}\,n^{\rm LC}_{\rm GJ}\,m_{\rm e} c^2})\simeq 7\times 10^5$}
(e.g., \cite{SironiSpitkovsky2014,cerutti2017,Hakobyan2019}).}.
For this to happen, the Larmor radius of such particles, $\rho_L \simeq 4\times10^{-3} \, R_{\rm LC}\, (\gamma^{\rm max}/ 7\times 10^7)$, 
should be larger than the size of the largest plasmoids.
Alternatively, cooled particles could be re-accelerated further in the CS layers and produce an IC emission in a second step.
In any case the phase coherence of the SR and IC pulses should be preserved, given the phase alignment of the HE and VHE  components in the data.
The formation of caustics for emission loci in the $1-2\, R_{\rm IC}$ region as shown in PIC studies
makes such scenarios plausible \cite{cps2016}.
The IC cooling time for particles with $\gamma^{\rm max} \sim  10^8$ scattering $\epsilon =0.1$~eV targets, 
$t_{\rm IC} \sim  50 \,{\rm ms}$,
is of the order of the pulsar period, and hence compatible with the above schemes.      
For generating the spectral model corresponding to this picture, we assume that the particles with energies beyond 
$\gamma^{\rm max}_{\rm SR}$ are from the same population as the 
one emitting SR, of which the energy spectrum is extended to beyond the 
SR cooling energy.       
The fit to the TeV data (shown for $\gamma^{\rm max} \sim  10^8$ and labeled $\rm{II_b}$ in Fig.~\ref{fig::spectral_models} of the main text)
results in an interaction volume $V^{\rm SR}_{\rm IC}\simeq 1.5 \times 10^{23}\,{\rm cm}^3$. 
This corresponds, as expected, to a larger volume than that estimated above, 
as the highest energy particles must have larger Larmor radii than those trapped and cooled through SR in the plasmoids.  
Various uncertainties, e.g. the correction factor $\upsilon\,\tau$ ($\zeta\,\omega_{\rm IR}\sim 1$, see above)
or the unknown lower bound of the target field energy range, make it difficult to go beyond this order-of-magnitude comparison.

The bulk movement of the striped wind, with Lorentz factor $\Gamma_{\rm w}$, can affect the physical picture  due to the Doppler 
boosting of radiations \cite{bogovalov2000,Aharonian2012,arkadubus2013,mochol15,mochol17,Tavernier2015, Jacob2019}.
In such a scheme, SR and IC take place in the wind co-moving frame 
where electrons are assumed to have an isotropic distribution. 
The observed emission is obtained after Doppler boosting of the quantities into the laboratory frame.
Assuming $B(\hat{r}) \sim  B_{\rm LC}/\hat{r}^2$ in the near wind region (e.g., \cite{arkadubus2013}),
and given the emission radius $\hat{r}=r/R_{\rm LC}$, $B^\prime(\hat{r}) = B(\hat{r})/\Gamma_{\rm w} $, 
and ${E^{\rm max}_{\rm SR}}^\prime= E^{\rm max}_{\rm SR}/2\,\Gamma_{\rm w} $, one can
rewrite (Eq. \ref{eq::SR}) as: ${\gamma^{\rm max}_{\rm SR}}^\prime\simeq 10^6 (\frac{E^{\rm max}_{\rm SR}}{{1.5\, \rm GeV}})^{1/2} \, \,\hat{r}$.
By requiring that the 20~TeV photons are produced by SR-emitting electrons, $\gamma^{\rm max}_{\rm IC} = 2\times \Gamma_{\rm w}\,  {\gamma^{\rm max}_{\rm SR}}^\prime$, one obtains a relation :\\
\begin{equation}
\Gamma_{\rm w} \simeq  22 \,\left(\frac{E^{\rm max}_{\rm SR}}{{1.5\, \rm GeV}}\right)^{-1/2} \, \left(\frac{E_{\rm VHE}}{{20\, \rm TeV}}\right)  \,\hat{r}^{-1}.
\end{equation}
The linear rise of the bulk Lorentz factor up to the fast magnetosonic point (e.g. \cite{cps2016}),  $\Gamma_{\rm w} = (1 + \hat{r}^2)^{1/2}$, provides 
 a second constraint, leading to a solution: $\Gamma_{\rm  w}\simeq \hat{r_{\rm e}}\simeq 5$.
Taking into account the exact shape of the GeV component modifies this solution somewhat, but the emission radius remains at 
a few $\, R_{\rm LC}$. As an example, a fit to the data with  $\Gamma_{\rm  w} =10$ at $\hat{r_{\rm e}}=5 $ 
is shown in Fig.~\ref{fig::spectral_models} in the main text (curve labeled $\rm{II_c}$).
A dissipation region at this distance is, however, not favored by PIC simulations which point to 
a SR emission region closer to the light cylinder (1 to 2~$\, R_{\rm LC}$).
An alternative boosted scenario, involving e.g. a re-acceleration of cooled particles, where SR and IC photons are emitted in separate zones could still be compatible with the above constraint if the formation of caustics provides the phase alignment of the two components.

\end{document}